\documentclass[fleqn,usenatbib]{mnras}

\usepackage{newtxtext,newtxmath}

\usepackage[T1]{fontenc}
\usepackage{ae,aecompl}

\usepackage{soul}
\usepackage{pdfpages}
\usepackage{subfig}
\usepackage{graphicx}	
\usepackage{amsmath}	
\usepackage{ulem}

\newenvironment{myfont}{\fontfamily{pag}\selectfont}{}

\newbool{show}
\boolfalse{show}
\newcommand\extra[1]{\ifbool{show}
    {\begin{myfont}\color{red}{#1}\end{myfont}}
    {}}
    
\newcommand\highlight{%
  \bgroup
  \expandafter\def\csname st\space\endcsname{\bgroup \ULdepth =-.8ex \ULset}%
  \markoverwith{\textcolor{yellow}{\rule[-.5ex]{.1pt}{2.5ex}}}%
  \ULon}
  
\newbool{showonline}
\booltrue{showonline}
\newcommand\online[1]{\ifbool{showonline}
    {#1}
    {}}

\makeatletter
\newcommand*{\addFileDependency}[1]{
  \typeout{(#1)}
  \@addtofilelist{#1}
  \IfFileExists{#1}{}{\typeout{No file #1.}}
}
\makeatother

\title[Galaxy scale jets]{A population of galaxy-scale jets discovered using LOFAR}

\author[B. Webster et al.]{
B. Webster$^{1}$\thanks{E-mail: brendan.webster@open.ac.uk},
J. H. Croston$^{1}$, B. Mingo$^{1}$, R. D. Baldi$^{2,3,4,5}$,B. Barkus$^{1}$, G. G\"urkan$^{6}$\newauthor M. J. Hardcastle$^{7}$, R. Morganti$^{8,9}$, H. J. A. R{\"o}ttgering$^{10}$, J. Sabater$^{11}$, T. W. Shimwell$^{8,10}$,\newauthor C. Tasse$^{12,13}$ and G. J. White$^{1,14}$
\\
$^{1}$School of Physical Sciences, The Open University, Walton Hall, Milton Keynes, MK7 6AA, UK\\
$^{2}$School of Physics and Astronomy, University of Southampton, Southampton, SO17 1BJ, UK\\
$^{3}$Dipartimento di Fisica, Università degli Studi di Torino, via Pietro Giuria 1, I-10125 Torino, Italy\\
$^{4}$INAF - Istituto di Astrofisica e Planetologia Spaziali, via Fosso del Cavaliere 100, I-00133 Roma, Italy\\
$^{5}$Istituto di Radioastronomia, INAF, via Gobetti 101, 40129, Bologna, Italy\\
$^{6}$CSIRO Astronomy and Space Science, PO Box 1130, Bentley WA 6102, Australia\\
$^{7}$Centre for Astrophysics Research, School of Physics, Astronomy and Mathematics, University of Hertfordshire, College Lane, Hatfield\\AL10 9AB, UK\\
$^{8}$ASTRON, Netherlands Institute for Radio Astronomy, Oude Hoogeveensedijk 4, Dwingeloo, 7991 PD, The Netherlands\\
$^{9}$Kapteyn Astronomical Institute, University of Groningen, P.O. Box 800, 9700 AV Groningen, The Netherlands\\
$^{10}$Leiden Observatory, Leiden University, PO Box 9513, 2300 RA Leiden, The Netherlands\\
$^{11}$SUPA, Institute for Astronomy, Royal Observatory, Blackford Hill, Edinburgh, EH9 3HJ, UK\\
$^{12}$GEPI \& USN, Observatoire de Paris, Université PSL, CNRS, 5 Place Jules Janssen, 92190 Meudon, France\\
$^{13}$Department of Physics \& Electronics, Rhodes University, PO Box 94, Grahamstown, 6140, South Africa\\
$^{14}$RAL Space, The Rutherford Appleton Laboratory, Chilton, Didcot OX11 0NL, UK
}

\date{Accepted XXX. Received YYY; in original form ZZZ}

\pubyear{2018}

\begin{document}
\label{firstpage}
\pagerange{\pageref{firstpage}--\pageref{lastpage}}
\maketitle

\begin{abstract}
The effects of feedback from high luminosity radio-loud AGN have been extensively discussed in the literature, but feedback from low-luminosity radio-loud AGN is less well understood. The advent of high sensitivity, high angular resolution, large field of view telescopes such as LOFAR is now allowing wide-area studies of such faint sources for the first time. Using the first data release of the LOFAR Two Metre Sky Survey (LoTSS) we report on our discovery of a population of 195 radio galaxies with 150 MHz luminosities between $3\times10^{22}$ and $1.5\times10^{25}\text{ W Hz}^{-1}$ and total radio emission no larger than 80 kpc. These objects, which we term galaxy-scale jets (GSJ), are small enough to be directly influencing the evolution of the host on galaxy scales. We report upon the typical host properties of our sample, finding that 9 per cent are hosted by spirals with the remainder being hosted by elliptical galaxies. Two of the spiral-hosted GSJ are highly unusual with low radio luminosities and FRII-like morphology. The host properties of our GSJ show that they are ordinary AGN observed at a stage in their life shortly after the radio emission has expanded beyond the central regions of the host. Based on our estimates, we find that about half of our GSJ have internal radio lobe energy within an order of magnitude of the ISM energy so that, even ignoring any possible shocks, GSJ are energetically capable of affecting the evolution of the host. The current sample of GSJ will grow in size with future releases of LoTSS and can also form the basis for further studies of feedback from low-luminosity radio sources.
\end{abstract}




\section{Introduction}
\label{sec:Introduction}

Throughout their history galaxies interact with their environments: accreting matter, forming new stars and evolving into the galaxies we see today \citep[for a detailed review  see][]{Benson2010GalaxyTheory}. To reproduce the observed numbers, sizes and distributions of galaxies, simulations such as Illustris \citep[][]{Weinberger2018SupermassiveSimulation} and EAGLE \citep[][]{Schaye2015TheEnvironments} must include some form of feedback that restricts the star formation rate (SFR). Active Galactic Nuclei (AGN) feedback from radio jets is one form of feedback, in which jets transport large amounts of energy into the surrounding environment, restricting the cooling rate. Suppression of cooling limits the rate at which material is accreted back into the galaxy where it can form stars and ultimately fuel the AGN itself \citep[e.g.][]{Hardcastle2020RadioJets, Fabian2012ObservationalFeedback, McNamara2007HeatingNuclei,Bower2006BreakingFormation,Croton2006TheGalaxies}. This type of feedback, sometimes referred to as `maintenance-mode' feedback, is typically associated with large, red galaxies and is believed to be the process limiting star formation to low rates in those systems.

The majority of the observational evidence for feedback from radio galaxies is associated with large jets of $\sim100 - 1000\text{ kpc}$ \citep[e.g.][]{Mullin2008Observed1.0}, capable of carrying energy far into the surrounding intracluster medium. There are also a growing number of studies looking into the effects of feedback from compact radio sources such as parsec-scale Gigahertz-Peaked Spectrum (GPS) and Compact Steep Spectrum (CSS) sources a few kpc in size \citep[e.g.][]{Bicknell2018RelativisticGalaxies, Tadhunter2016TheObservations}, along with some studies of intermediate-size radio structures \citep[e.g.][]{Jarvis2019PrevalenceQuasars, Jimenez-Gallardo2019COMP2CAT:Universe}. However, the advent of telescopes such as the International LOw Frequency ARray (LOFAR) Telescope, with its combination of high sensitivity to both compact and extended emission \citep[][]{Shimwell2017TheRelease}, allows for the systematic identification of sources with sizes on scales similar to that of the host galaxy, together with the study of the potential effects these sources can have upon their host environments. Unlike the FR0 class of unresolved objects \citep[][]{Baldi2015ARadio-galaxies,Baldi2018FR0Galaxies}, these are resolved sources with radio emission a few tens of kpc in size, making them bigger than the majority of GPS/CSS sources with jets that are large enough to have escaped the dense environment at the core of the host galaxy. However, unlike larger radio galaxies, they are still small enough to be interacting with, or have recently impacted, a substantial portion of the host's interstellar medium (ISM), directly affecting the evolution of the host. We refer to these radio sources as galaxy-scale jets (GSJ). 

Their sizes make GSJ ideal for directly studying the role and importance of feedback from radio jets. The importance of understanding feedback from these sources has been highlighted by the discovery that some GSJ are capable of creating jets that are strong enough to develop shock fronts that heat their environment \citep[e.g.][]{Croston2009High-energyA, Mingo2011MarkarianSeyfert,Hota2012NGCFeedback}. We also know that these shocks can travel into the host galaxy and that they can carry energy equivalent to the thermal energy of the ISM \citep[][]{Croston2007Shock3801}. This strongly suggests the possibility that, like supernova shocks, they can affect star formation within the host. This influence could either take the form of positive feedback, in which gas within the host is compressed leading to an increase in star formation \citep[][]{ Markakis2018High-resolutionFormation, Salome2017InefficientA, Kalfountzou2014Herschel-ATLAS:Quasars} or negative feedback, in which gas within the host is heated leading to a reduced SFR \citep[][]{Gurkan2015iHerschel/iGalaxies,Guillard2015ExceptionalGalaxy}. GSJ may also be an intermediate stage in the potential evolution of the smaller GPS/CSS sources into larger FRI/FRII sources \citep[][]{Fanaroff1974TheLuminosity}. 

At present very little is known about the galaxies that host GSJ. The majority of previously discovered GSJ are hosted by ellipticals but a small number are hosted by spiral galaxies \citep[e.g.][]{Hota2006RadioNGC6764, Gallimore2006ANuclei, Croston2008iChandra/i6764, Mingo2012ShocksGalaxy}. It is therefore presently unclear whether these smaller jets form part of the evolutionary sequence of a `typical' radio-galaxy or a separate population. It is also not yet known how GSJ and the population of large-scale double-lobed radio galaxies hosted by spiral galaxies, the so-called spiral DRAGNs \citep{Mulcahy2016TheDRAGN}, may be related.

In order to investigate the importance of GSJ in shaping galaxy evolution we use the first data release (DR1) of the LOFAR Two Metre Sky Survey \citep[LoTSS;][]{Shimwell2019TheRelease,Williams2019TheSurvey} to find a sample that is large enough to draw statistical conclusions. LoTSS is a radio survey undertaken at $150\text{ MHz}$ using the International LOFAR Telescope \citep[for a full description of LOFAR see][]{vanHaarlem2013LOFAR:ARray}. LoTSS DR1 covers the \textit{Hobby-Eberly} Telescope Dark Energy Experiment (HETDEX) Spring field, an area of sky $424 \text{ deg}^2$ in size between right ascension $161.25^{\circ}$ - $232.5^{\circ}$ and declination $45^{\circ}$ - $57^{\circ}$. The LoTSS survey is about an order of magnitude more sensitive than the Faint Images of the Radio Sky at Twenty Centimetres (FIRST) \citep[][]{Becker1995TheCentimeters} or the NRAO VLA Sky Survey (NVSS) \citep[][]{Condon1998TheSurvey} for sources with a typical spectral index of $\alpha= -0.7$ (where $S_v \propto v^{\alpha}$). LOFAR's combination of short and long baselines make it sensitive to both compact and extended emission, avoiding the need to combine catalogues such as NVSS and FIRST.

DR1 of the LoTSS catalogue contains over 300,000 sources and covers $\sim 2$ per cent of the planned sky area \citep[][]{Shimwell2019TheRelease}. The sheer number of sources make it impossible to visually search through the catalogue for GSJ. We therefore devised a system that uses the host and radio morphology to reduce the number of potential sources to more manageable numbers. We then used a combination of size criteria, existing AGN/star-formation separation methods, and visual inspection to identify our sample. These methods have the advantage that they can be easily implemented in future catalogue releases. We then studied the properties of our sample to determine how common these objects are, how they relate to larger radio galaxies, and their potential energetic impact on the host galaxies.

This paper is structured as follows. Section~\ref{sec:TheData} describes the method used to find our sample of GSJ. Section~\ref{sec:Results} describes the radio and host properties of our GSJ as well as the environments in which they are typically found. Section~\ref{sec:Results-Fraction} describes how common GSJ are when compared to both the overall population of galaxies as well as the wider population of AGN. Section~\ref{sec:Results-Energetics} looks at the potential impact of GSJ upon their hosts. Section~\ref{sec:Discussion} places our results within the wider context of galaxy evolution whilst Section~\ref{sec:Conclusions} summarises our findings. Throughout this paper we assume cosmological parameters of $\Omega_m=0.3$, $\Omega_{\Lambda} = 0.7$ and $H_0 = 70 \text{ km s}^{-1} \text{ Mpc}^{-1}$. For the spectral index, $\alpha$, we use the definition of radio flux density $\text{S}_{\nu}\propto \nu^{\alpha}$.

\section{Sample Selection}
\label{sec:TheData}

We want to generate a large, reliable sample of radio galaxies with galaxy-scale jets in an automated way using readily-available catalogue data, so that it can be applied to larger sky areas in the future. In Section~\ref{sec:TheData-AS} we discuss our approach to generating an automated sample of GSJ with Table~\ref{tab:sample_sizes} showing the sample size at each step of the process. In order to validate our method and ensure that our selection criteria are not introducing any biases (e.g. against a particular type of host), we also generate a smaller, visually selected, clean sample that can be used for more detailed investigations of the properties and impact of GSJ. Our approach to selecting this clean sample is discussed in Section~\ref{sec:TheData-VS}.

The starting point for our sample selection was the LoTSS DR1 value-added catalogue, of which we consider only those 231,716 sources (about 70 per cent of the total) with an optically identified host \citep[for the full catalogue description see][]{Williams2019TheSurvey}. As explained in Sections~\ref{sec:ratio-selec} and \ref{sec:TheData-AS}, we also made use of the \citet{Hardcastle2019Radio-loudSources} catalogue of 23,444 radio-loud AGN selected from the DR1 catalogue, based on a combination of radio excess, spectroscopic and infrared colour diagnostics. However, we wanted to avoid biases from pre-judging the likely host-galaxy colours and emission line properties for this population, where the radio excess may be relatively low, and so do not use the AGN catalogue as our initial starting point. Our selection criteria, which are explained in detail below and summarized in Table~\ref{tab:sample_sizes}, include (i) improved size estimation and a size cut-off, (ii) a threshold in the ratio of radio to host galaxy size, as measured along the jet axis, (iii) AGN/star-formation separation based on the catalogue of \citet{Hardcastle2019Radio-loudSources}. 

\begin{table}
    \centering
    \begin{tabular}{l|c}
        \hline
         Selection step & Sample size\\
         \hline
         DR1 sample with optical IDs&231,716\\
         Resolved with LoMorph size measurement&15,472\\
         Total length less than 80 kpc&2,987\\
         Jet:Galaxy ratio in range 2-5&454\\
         Automatic Sample (AS)&192 (167{\scriptsize(s)}, 25{\scriptsize(p)})\\
         Visual Sample (VS)&52 (48{\scriptsize(s)}, 4{\scriptsize(p)})\\
         Total Sample (TS)&195 (170{\scriptsize(s)},  25{\scriptsize(p)})\\
         \hline
    \end{tabular}
    \caption{The size of the sample at each step of the selection process. The AS, VS and TS show how many of each sample have spectroscopic (s) and how many have only photometric (p) measurements.}
    \label{tab:sample_sizes}
\end{table}

\subsection{Size selection}
\label{sec:size-selec}

The LoTSS catalogue includes source size estimates determined by the Python Blob Detector and Source Finder {\sc pyBDSF} \citep[][]{Mohan2015PyBDSF:Finder}. However, these ellipse regions were obtained through Gaussian approximations to the source extent, and sometimes do not accurately represent the shape or size of the radio emission. For small objects that may have complex brightness distributions, which are of particular interest for this work, pyBDSF tends to underestimate their true size \citep[see][for details]{Mingo2019RevisitingLoTSS}. We therefore used part of the LoMorph code of \citet{Mingo2019RevisitingLoTSS} to measure the total extent of the radio emission above a threshold set at 5 times the RMS noise. We first eliminated unresolved sources using the criteria of \citet{Shimwell2019TheRelease}, and rejected those sources with a measured flux below $2\text{ mJy}$ as too faint to measure. This left 25,128 sources, of which 15,472 had sufficient flux for LoMorph to determine an accurate size.

A size cut-off was then applied to reduce our sample to objects consistent with being GSJs, i.e. with radio emission on scales comparable to their host. Typical large elliptical galaxies have half-light radii up to approximately 20 kpc, though the full extent of the galaxy will be significantly larger \citep[][]{Forbes2017TheImaging}. We therefore chose to exclude all sources with a total extent greater than 80 kpc, in other words larger than twice the typical galaxy half-light diameter. Projection effects may mean that some of these sources are in fact larger than 80 kpc, however, since we do not know inclinations we make no allowances for this in our calculations. Applying this size criterion left 2,987 sources.

\subsection{Jet:galaxy ratio}
\label{sec:ratio-selec}

To further refine the sample to include only objects with jet-related emission on a similar scale to the host, we divided the size of the radio emission by the optical size of the host to get the jet:galaxy (size) ratio for all 2,987 sources. Since jets can be found at a wide range of angles relative to the host's major and minor axes \citep[e.g.][]{Kharb2016A}, for example see Figure~\ref{fig:DiffPositionAngles}, we measured the size of the host along a line defined by the position angle of the radio emission (taken from LoMorph). This avoided prejudicing our selection against highly elongated hosts.

\begin{figure}
    \centering
    \includegraphics[width=0.45\textwidth]{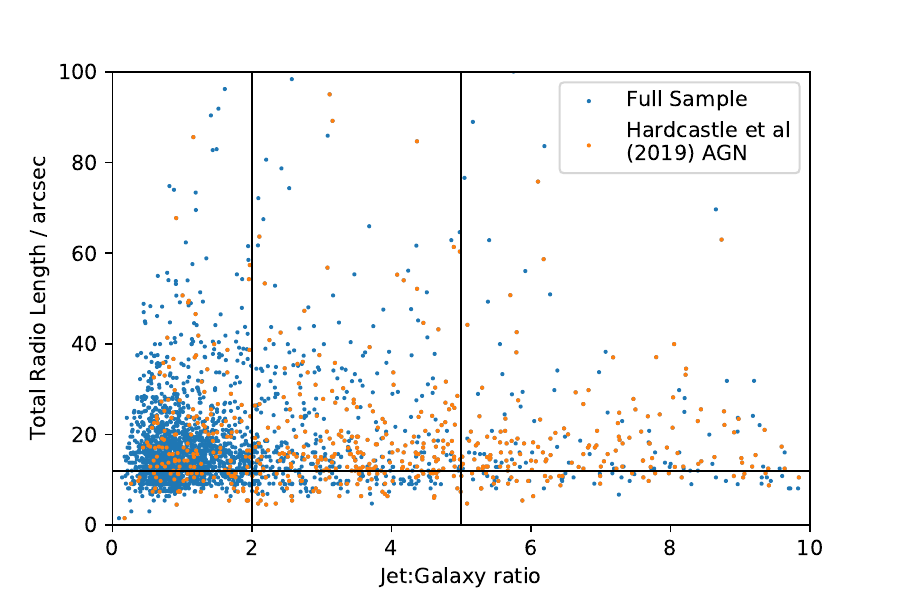}
    \caption{Plot of the total angular size of the radio emission versus the jet:galaxy ratio. The vertical lines and horizontal lines represent the jet:galaxy ratios and angular sizes used to reduce the size of the candidate sample. The objects within the LoTSS DR1 sample and the subset identified as radio-loud AGN by \citet{Hardcastle2019Radio-loudSources} are shown}
    \label{fig:GSJ_Selection_Ratio}
\end{figure}

The host galaxy sizes were obtained from the SDSS DR14 catalogue \citep[][]{Abolfathi2018TheExperiment}. A further 52 sources were excluded due to not having an SDSS match within 3 arcsec. The catalogued i-band deVaucouleurs radius and ellipticity were used to determine the host angular size along the axis of the radio emission, except where the i-band radius was $>1.5\sigma$ from the mean across all five bands, in which case another band close to the mean was used. The LoMorph-measured radio size was divided by this host radius to obtain a jet:galaxy ratio for each object. Figure~\ref{fig:GSJ_Selection_Ratio} shows the distribution of jet:galaxy ratio vs angular size, where a jet:galaxy ratio $>1$ indicates emission extending beyond the host galaxy effective radius.

We next used visual inspection to determine suitable automated thresholds in angular size and jet:galaxy ratio. At angular sizes less than twice the LoTSS resolution ($12$ arcsec), it became difficult to be certain that any extended structure is genuine and not due to calibration uncertainties, and so we eliminated all sources with angular size $<12$ arcsec, leaving 2,105 sources. 

A lower threshold in jet:galaxy ratio is needed to reduce contamination from star-forming galaxies, while an upper threshold is needed to exclude objects where the majority of energy must be deposited at a large distance from the host. To determine the lower threshold, we examined the subset of our candidates with jet:galaxy ratio between 1 and 3. In the range 1 -- 2, the radio emission from 35 of 76 candidates has the appearance of a star-forming galaxy, whereas in the range 2 --3, only 4 of 58 candidates appear dominated by radio emission from star formation. We therefore adopted a lower jet:galaxy ratio cut-off of 2, as the best compromise to enable fully automated selection without a high level of contamination. Unfortunately this unavoidably excludes the smallest GSJs whose jets are well embedded within the central regions of the galaxy. Future higher-resolution studies (e.g. future releases of LoTSS using the international baselines), should be able to resolve jet-related structures on sub-galactic scales, allowing these sources to be identified using our methods.

\begin{figure}
\centering
\includegraphics[width=0.9\columnwidth, trim={55 10 55 25}, clip=true]{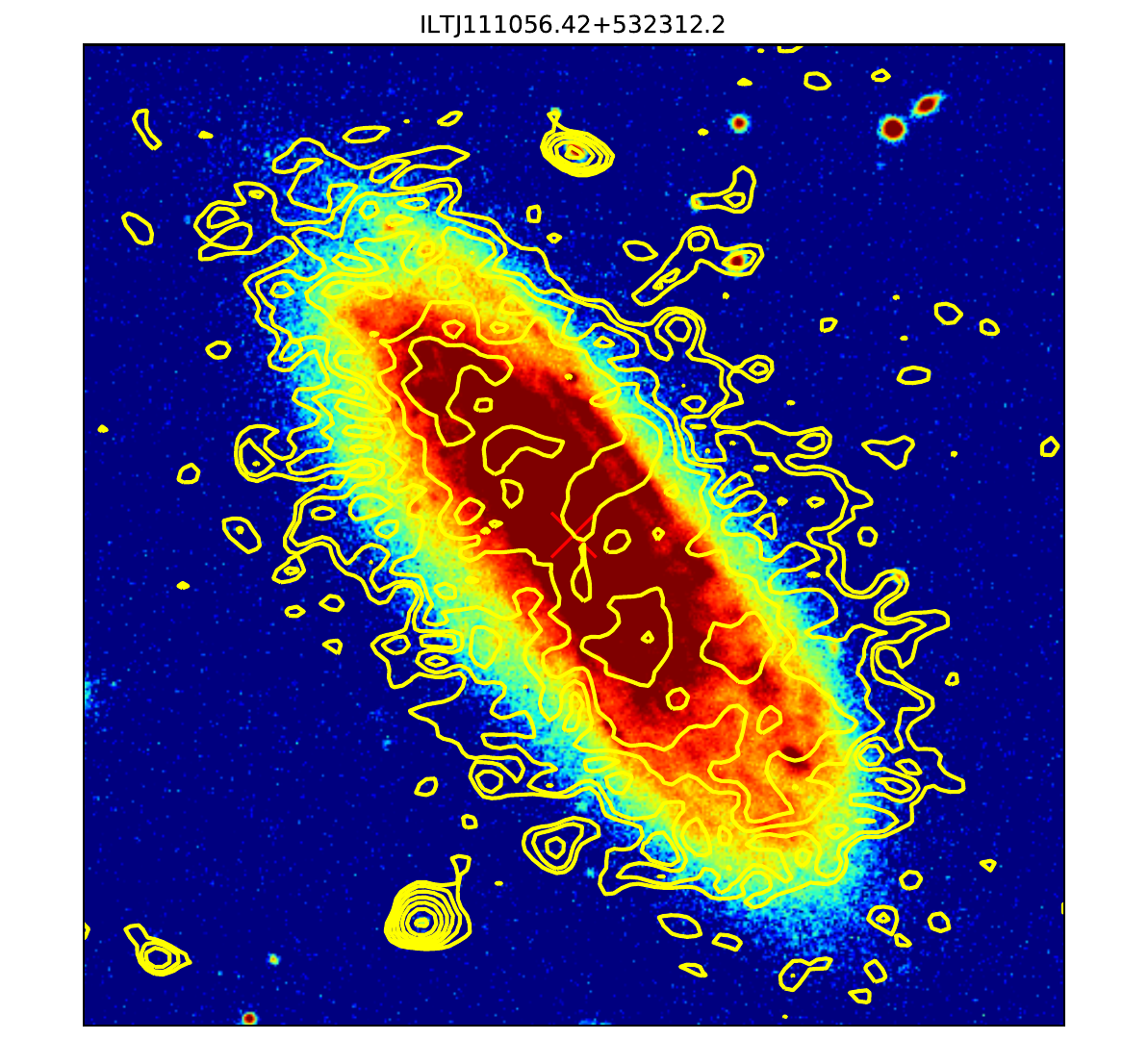}
\caption{Image of ILTJ111056.42+532312.2. An example of an AS galaxy whose radio emission appears to be related to star formation rather than AGN activity.}
\label{fig:ILTJ111056}
\end{figure}

A further compromise is needed in choosing an upper threshold in jet:galaxy ratio. Sources such as spirals and highly elongated ellipticals with jets closer in projection to the minor axis could have high jet:galaxy ratios whilst still having jets that are small enough to be directly influencing the host's evolution (e.g. in a way similar to the jet-induced turbulence and outflowing ionised gas bubbles within the host's ISM proposed by \citet{Jarvis2019PrevalenceQuasars}. A jet:galaxy ratio of 5 was chosen as the upper cut-off.

Applying the lower and upper jet:galaxy ratio cut-offs leads to a candidate sample of 454 sources, which are used to define an automatically selected sample, and a smaller, visually selected sample below.

\subsection{The Automatically Selected Sample}
\label{sec:TheData-AS}

Of the 454 candidate sources obtained by applying the size and jet:galaxy ratio thresholds, the 192 that are classified as AGN by \citet{Hardcastle2019Radio-loudSources} form our `Automated Sample' (AS). The methods of Hardcastle et al. provide a straightforward automated method to avoid significant contamination by radio sources dominated by star formation. Based on our visual check of the AS there does appear to be some low level of contamination, with about $2$ per cent potentially being misclassified star-forming galaxies (discussed further in Section~\ref{sec:Results-Hosts-Morphology}), plus an additional $6$ per cent where the radio emission seen by LOFAR from the GSJ is potentially blended with another source. For these sources it is possible that the catalogued radio fluxes may be slightly too high. It is also possible that some of the catalogued fluxes include an element of non-jetted AGN-related radio emission. However, within this paper we make no modifications to catalogued flux density values.

As with any flux limited survey, some of our sources may have additional extended emission below the surface brightness limitations of LoTSS. Any such sources would be correspondingly larger and might therefore not qualify as GSJ according to our criteria. Even if this were the case for some of our sources, at least some of the energy associated with the emission seen by LoTSS must be transferred locally, so that these sources could still be having an impact on galaxy scales. In the future, higher-sensitivity LoTSS deep fields data could give an estimate of what percentage of GSJ have faint extended emission on larger scales. As above, we make no allowances for these potential contaminants within this paper.

\subsection{The Visually Selected Sample}
\label{sec:TheData-VS}

\begin{figure*}
\centering
\begin{tabular}{ccc}
\includegraphics[width=5.0cm, trim={55 10 55 25},  clip=true]{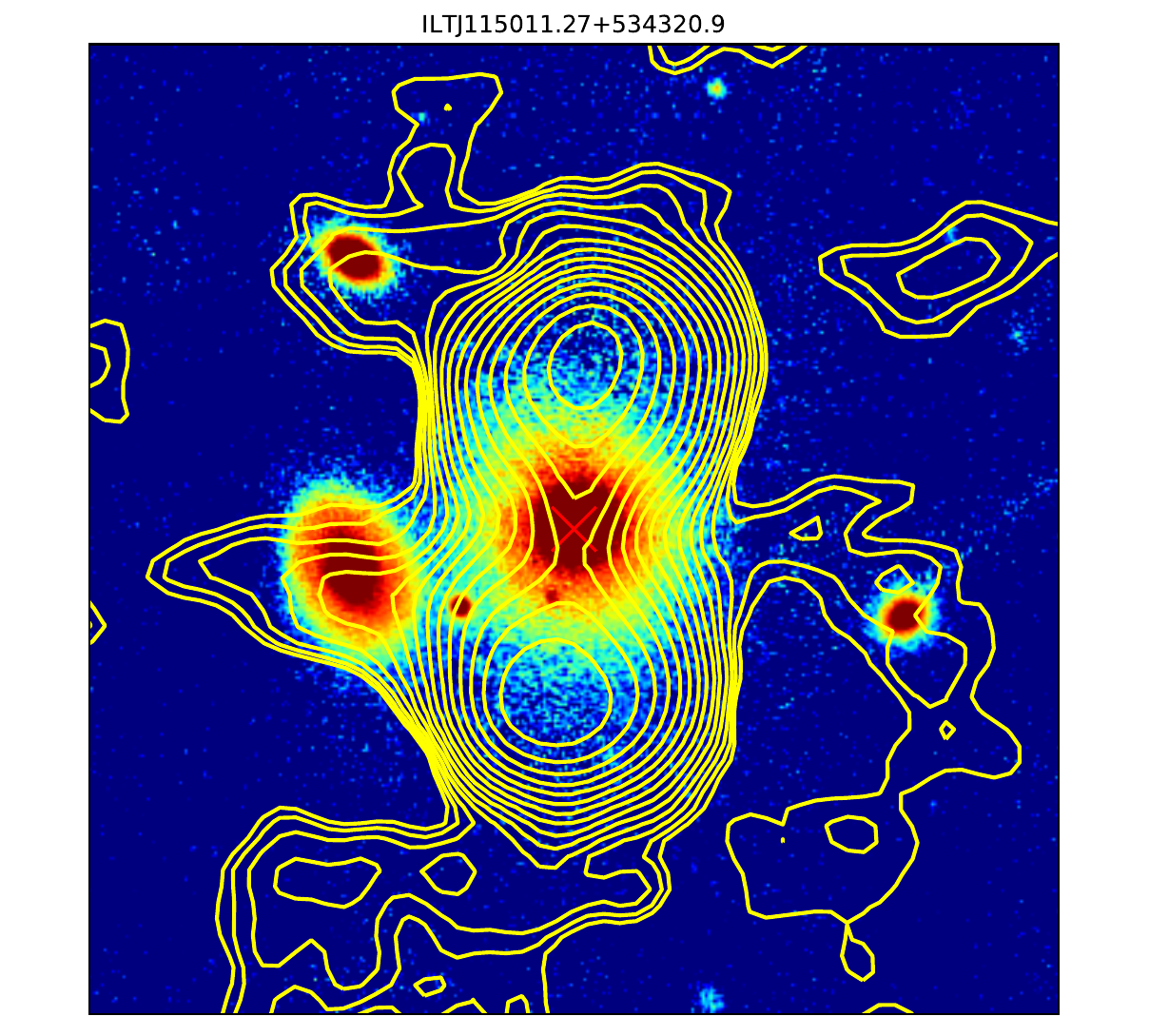} &
\includegraphics[width = 5.0cm, trim={55 10 55 25}, clip=true]{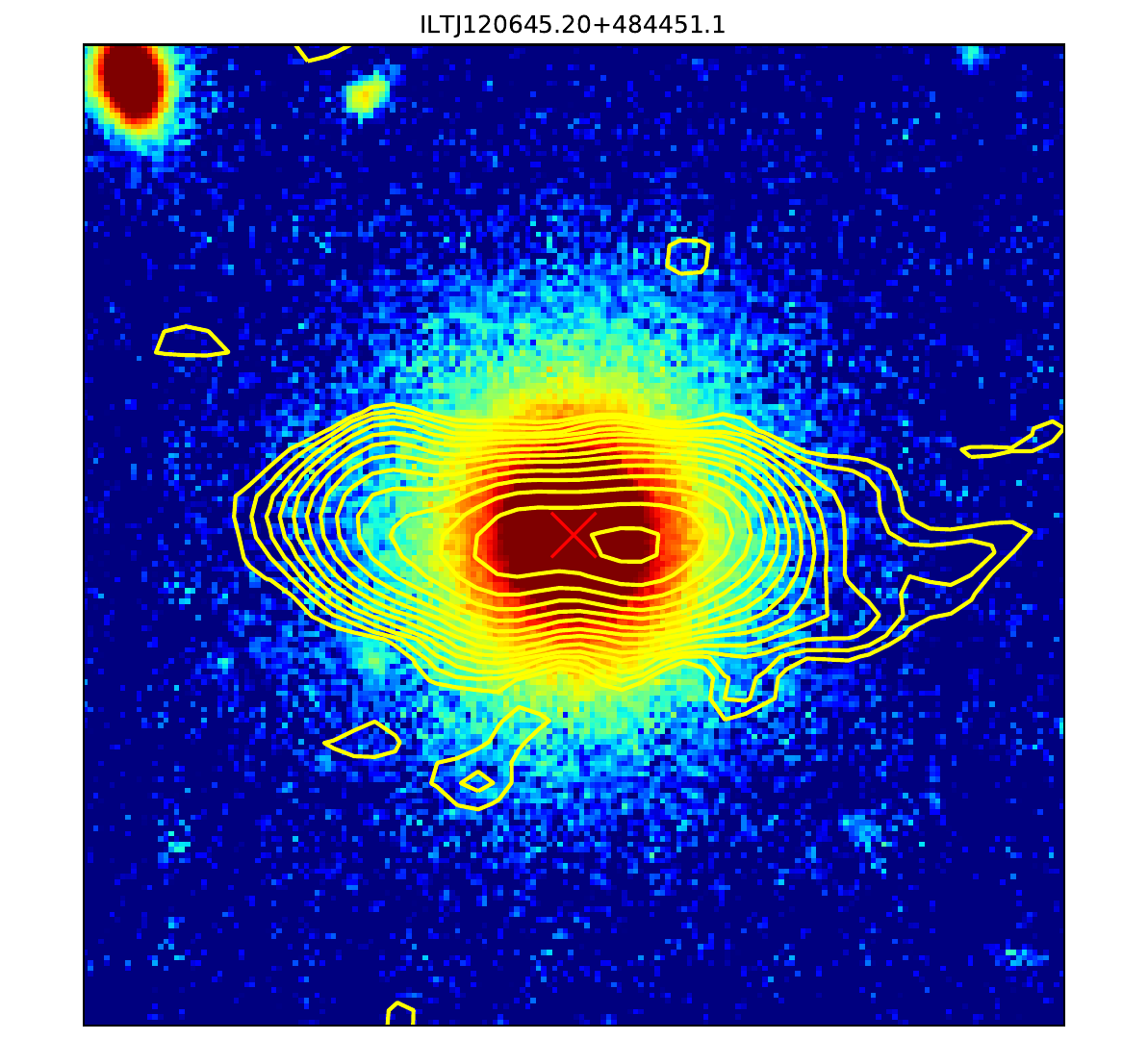} &
\includegraphics[width=5.0cm, trim={55 10 55 25}, clip=true]{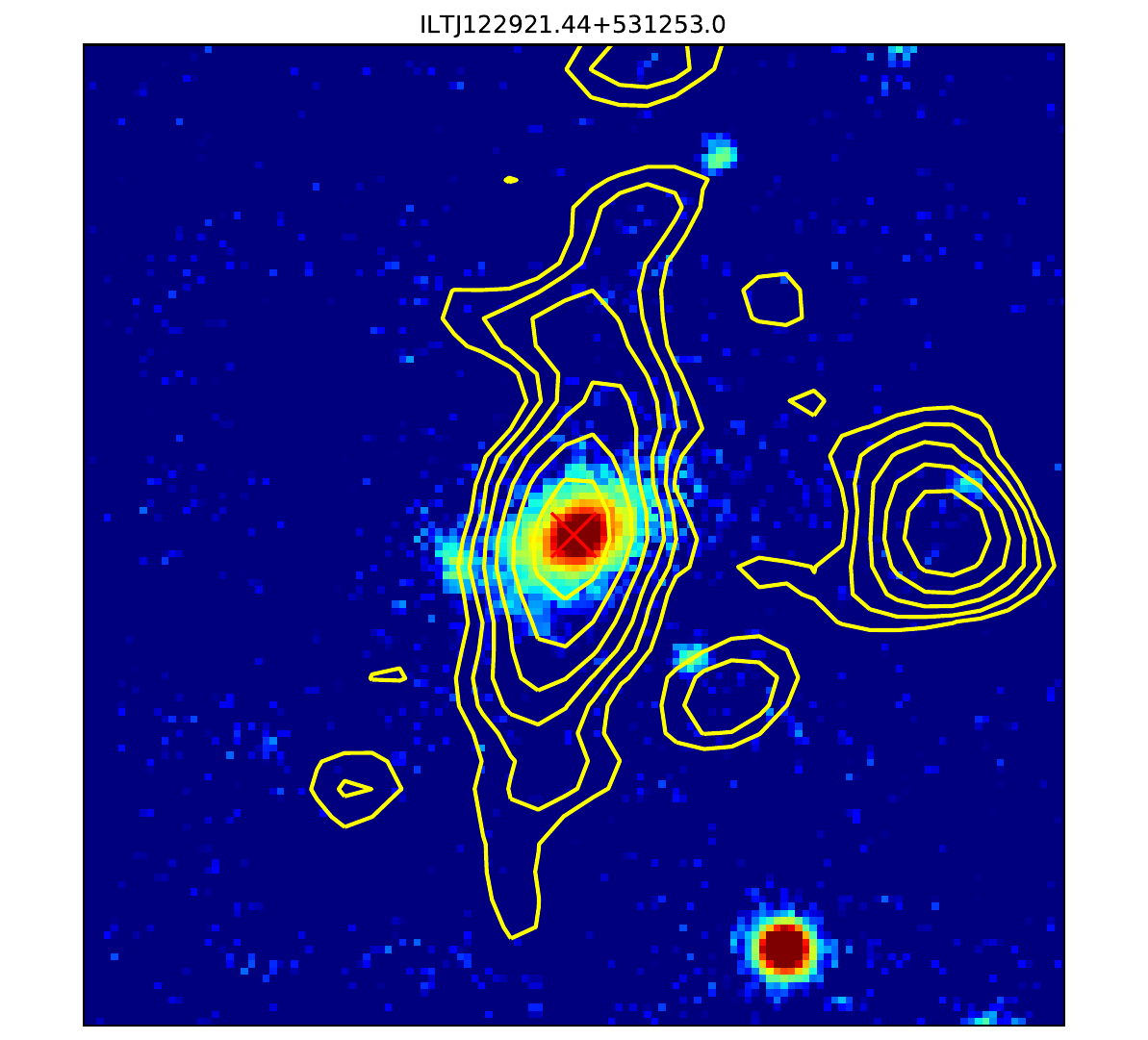} \\
\includegraphics[width=5.0cm, trim={55 10 55 25},  clip=true]{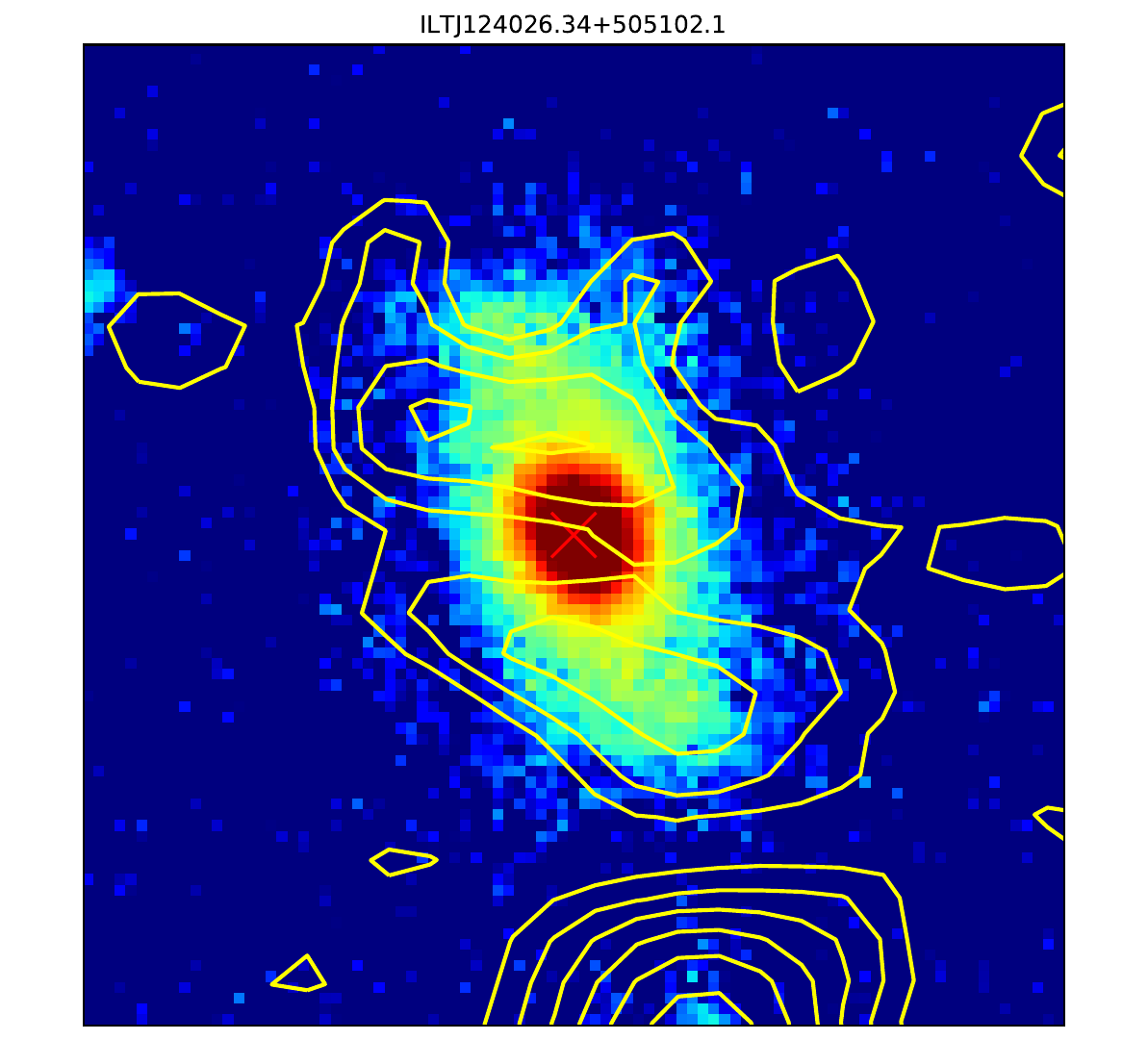} &
\includegraphics[width=5.0cm, trim={55 10 55 25},  clip=true]{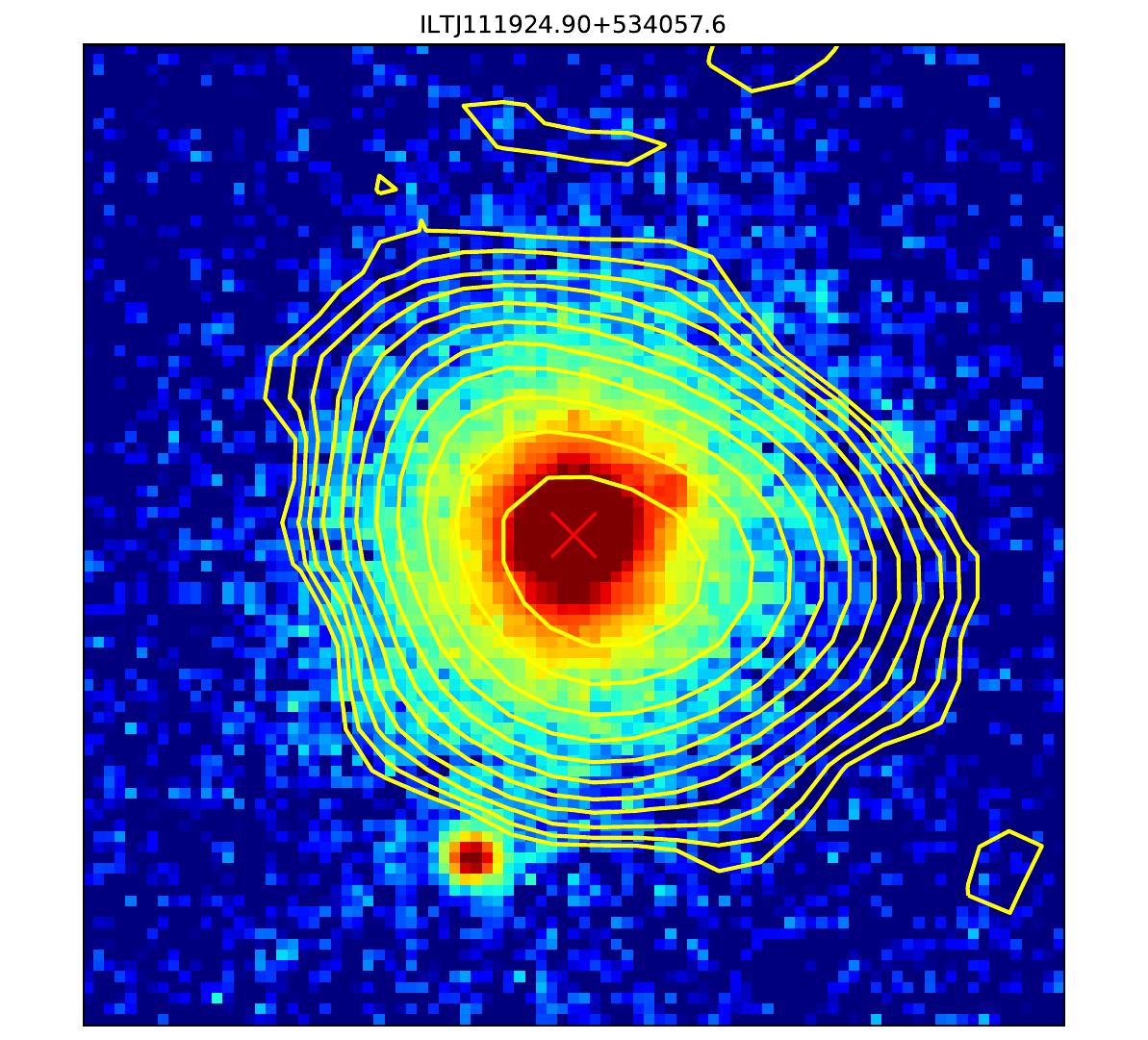} &
\includegraphics[width=5.0cm, trim={55 10 55 25},  clip=true]{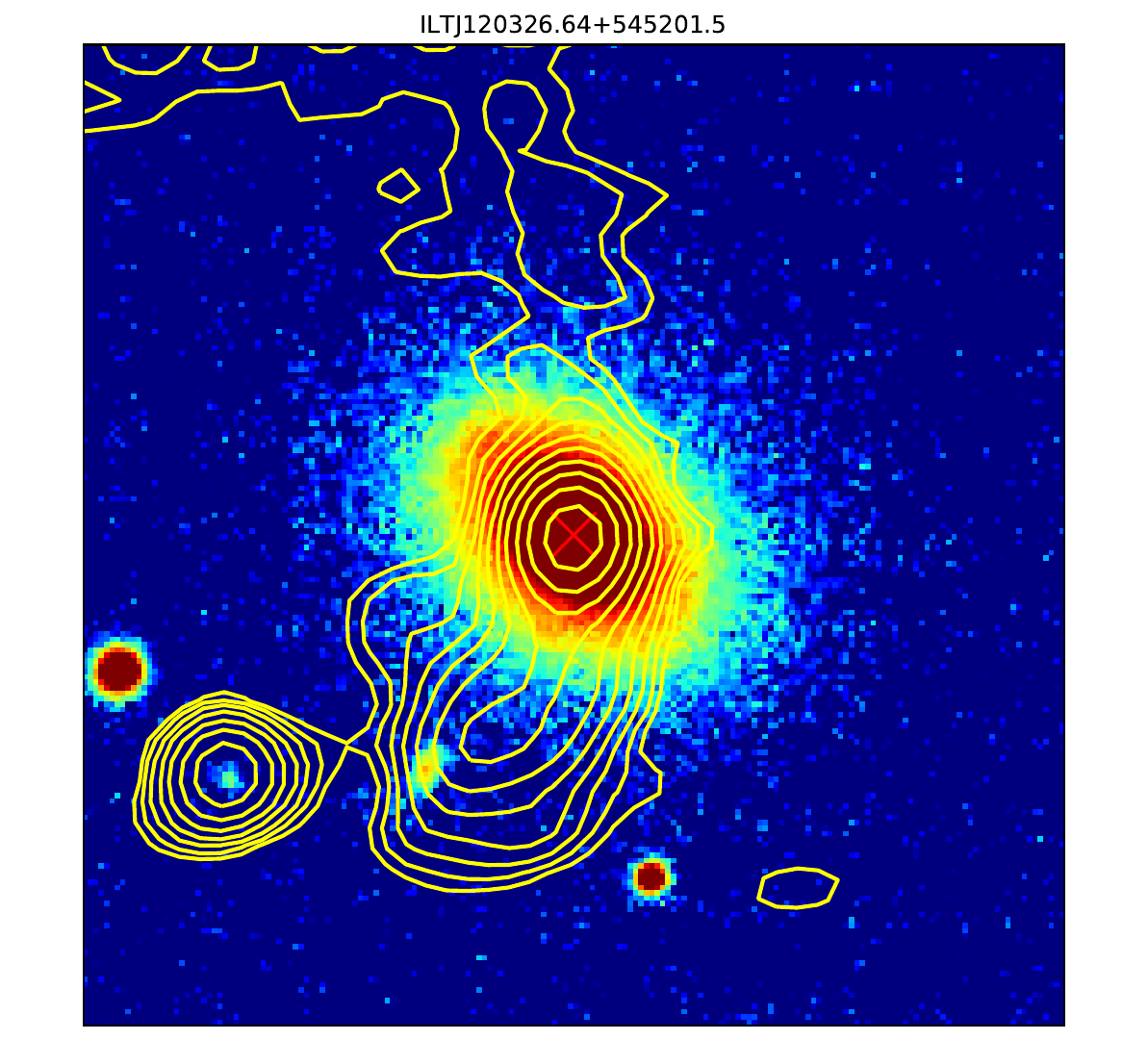} 
\end{tabular}
\caption{The top row shows examples of sources from the VS, whilst the bottom row has examples of AS sources. The leftmost column shows the difference in source structure, the middle shows the difference in ellipticity and the rightmost column shows the difference in symmetry between the samples.}
\label{fig:ASS_VSS_Comparison}
\end{figure*}

In order to verify the selection methods used to find the AS, we visually inspected all 454 candidate sources (Section~\ref{sec:ratio-selec}) to identify those sources with unambiguous jetted structure. Unlike the AS which aims to be as complete as possible, this sample, which we refer to this as our `Visual Sample' (VS), is intended to be as clean as possible. The VS can therefore also be used for detailed investigations and for optimising follow-up observations. When inspecting the sources we applied the following procedure:
\begin{itemize}
    \item Sources with a clear double-lobed morphology were always considered as GSJ. For example, in the leftmost column of Figure~\ref{fig:ASS_VSS_Comparison} the GSJ (top) has two clearly defined, roughly circular, radio features at the opposite ends of the jet. In contrast the rejected source (bottom) has quite diffuse emission with two poorly defined circular features buried within the radio emission. Whilst these features may be due to AGN activity it could also be caused by star-forming regions within the host galaxy.
    \item The shape of the radio emission. Sources with circular emission or a brightness distribution closely matching that of the host could easily be caused by star-forming activities and were rejected. In contrast strongly elliptical radio structures with large amounts of flux and aspect ratios greater than about two were typically considered to be GSJ. For example, the middle column of Figure~\ref{fig:ASS_VSS_Comparison} shows a GSJ source (top) with much more pronounced ellipticity than the rejected source (bottom).
    \item Sources where there was a strong asymmetry in the radio emission on either side of the host. Whilst most of these are still likely to be jetted sources, the asymmetry may indicate that either the jets are inclined at a significant angle compared to the plane of the sky so that the source is not a true GSJ, that some of the flux is attributable to a secondary background source or that the host has been incorrectly identified. For example, in the rightmost column of Figure~\ref{fig:ASS_VSS_Comparison} the radio emission on either side of the GSJ (top) is very similar whilst the radio emission on one side of the rejected source (bottom) is much longer than the other suggesting that the jets may be bent/inclined towards us and that this is not a true GSJ.
\end{itemize}

This inspection resulted in a sample of 52 GSJ which form the VS, of which 49 are also in the AS. The three sources in the VS that are not in the AS are ILTJ112543.06+553112.4 (hereafter ILTJ112543), ILTJ121847.41+520128.4 (hereafter ILTJ121847) and ILTJ123158.50+462509.9. The first two appear to be spiral hosted radio galaxies which, due to their unusual hosts were not included in the Hardcastle et al. catalogue, whilst the third is an elliptical host with an unusually low luminosity compared to its star formation rate. All three sources are discussed in detail in  Appendix~\ref{sec:AppendixAdditionalVSSSources}. It therefore appears that when combined with our selection techniques, the methods of \citet{Hardcastle2019Radio-loudSources} can be used to find GSJ, although these three sources indicate that this will exclude a small percentage ($\sim1.5$ per cent of our sample) of genuine GSJ.

The 143 sources in the AS that are not in the VS predominantly have either one-sided or round-ish emission. Whilst the radio excess used by \citet{Hardcastle2019Radio-loudSources} strongly suggests that these sources are radio galaxies, their morphology is too ambiguous to be included in our visual sample.

\begin{figure}
    \centering
    \includegraphics[width=\columnwidth]{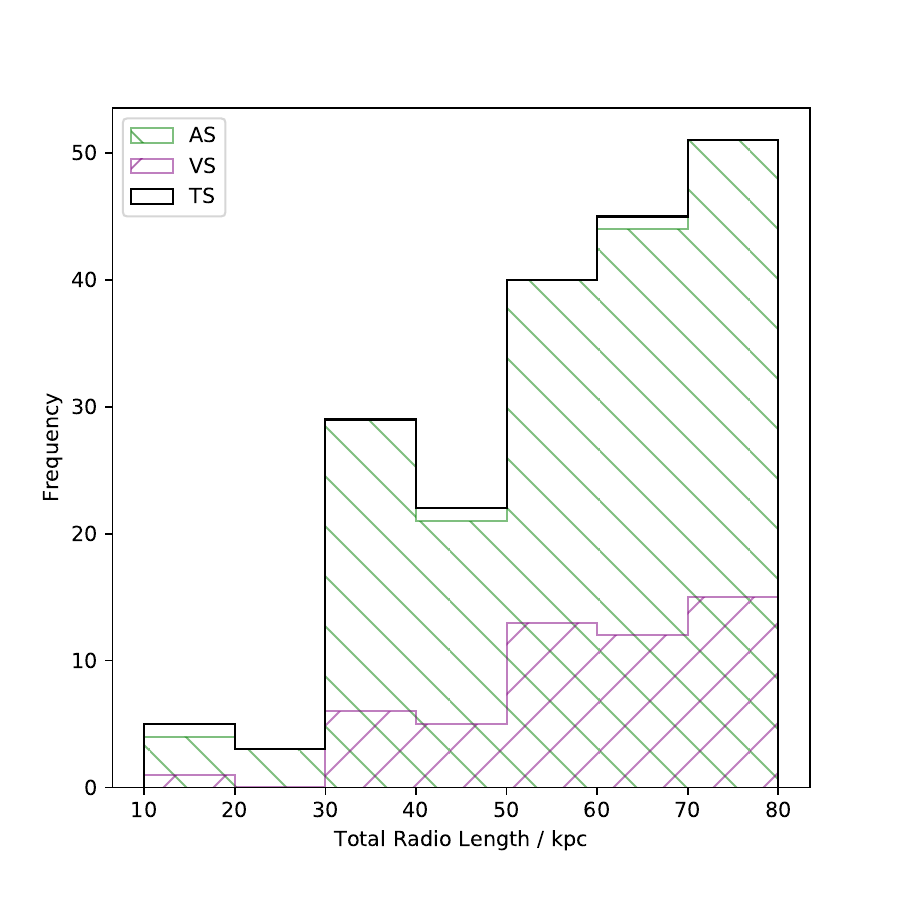}
    \caption{Distribution of total radio emission lengths for the TS, AS and VS.}
    \label{fig:JetSizes}
\end{figure}

In order to have as large a sample as possible, we combine all the unique sources in both the AS and VS to produce the 195 sources that form our `Total Sample' (TS). A summary of the three samples is given in Table~\ref{tab:sample_sizes} and the two-sided jet lengths are shown in Figure~\ref{fig:JetSizes}.

\section{Results}
\label{sec:Results}

We wish to compare the properties of our GSJ samples to those of larger radio galaxies, in order to discover how they fit into the overall picture of galaxy life cycles and the potential effects of feedback on the host. In the first subsection we discuss the redshift, morphology and luminosity distributions of the GSJ radio emission. In the second subsection we investigate the host properties of our sample to see what type of galaxies host GSJ. In order to compare the radio properties of our GSJ with a parent sample of larger radio galaxies, we use the 3,820 resolved objects with a redshift less than 0.5 from the catalogue of \citet{Hardcastle2019Radio-loudSources}, referred to within this section as H19.

\subsection{Redshift and Radio Properties}
\label{sec:Results-RadioProperties}

\begin{figure*}
\centering
\begin{tabular}{cc}
\includegraphics[width=0.45\textwidth]{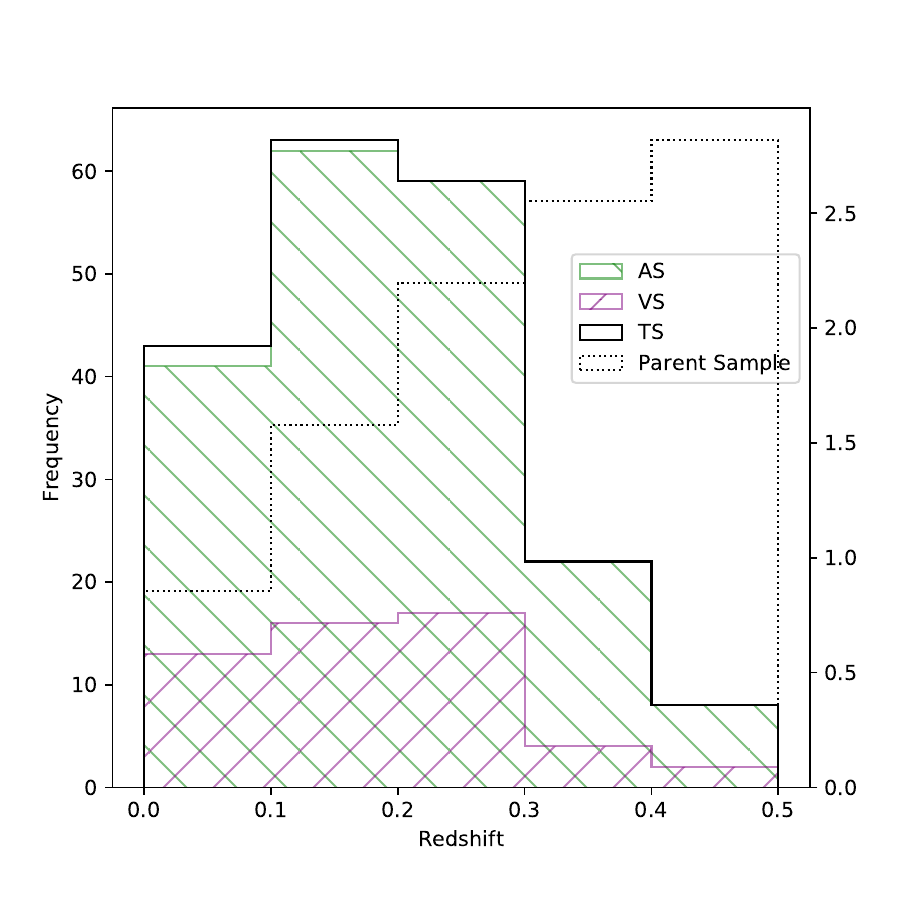} &
\includegraphics[width=0.45\textwidth]{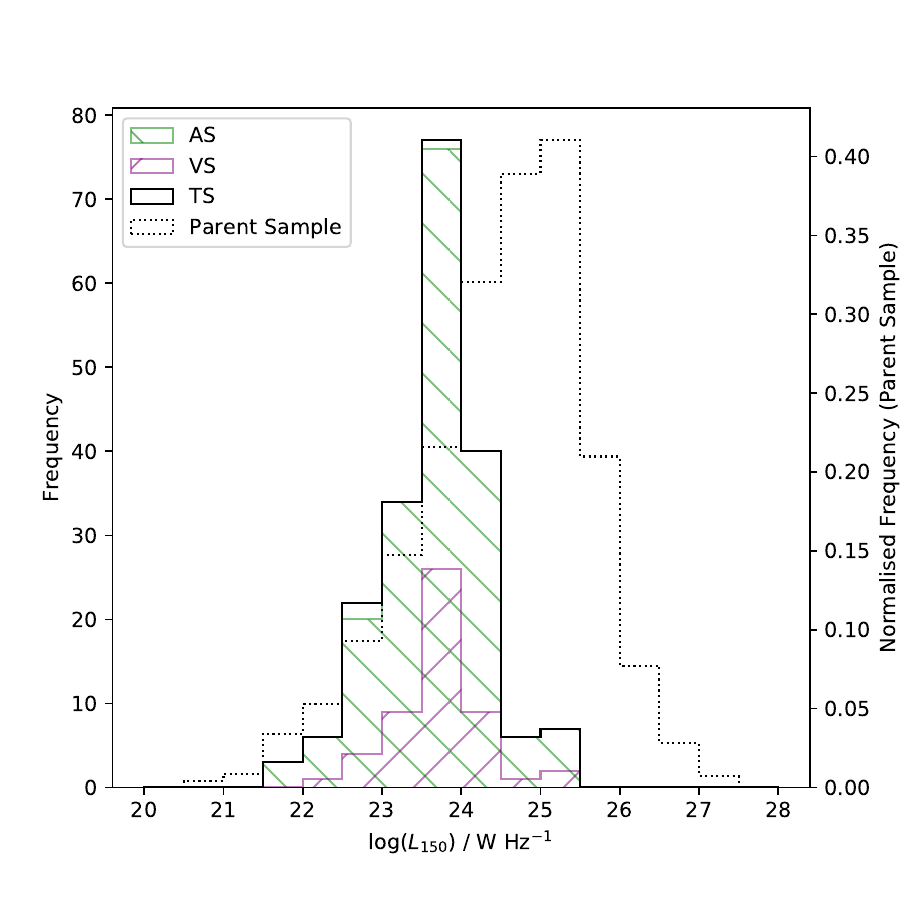}
\end{tabular}
\caption{Redshift (left) and luminosity (right) distributions for the AS, VS and TS. The dotted black line in each diagram shows the normalised distribution for the H19 parent sample.}
\label{fig:RadioProperties}
\end{figure*}

The distributions of redshift and luminosity for our sample are shown in Figure~\ref{fig:RadioProperties}. As redshift increases it is increasingly difficult to identify GSJ. Hence, it is to be expected that the distribution of our sample is biased towards lower redshifts, with a marked decrease in the numbers of objects after 0.3. The luminosities of the TS are tightly distributed about a mean value of $\text{log}(L_{150})=23.7$, with a standard deviation of 0.6. The AS and VS have similar distributions. These values are offset from the H19 sample whose mean luminosity is $\text{log}(L_{150})=24.5$ and exhibits a wider spread of values, with a standard deviation of 1.1. Our population of GSJ therefore forms a distinct subset within the H19 sample, as expected given our selection criteria. 

It is also informative to compare our GSJ with the catalogue of \citet{Jimenez-Gallardo2019COMP2CAT:Universe}, hereafter JG19, which contains 43 FRII objects of similar physical size to our GSJ, but with 150 MHz luminosities in the range $\sim10^{24}$ to $10^{26}\text{ W Hz}^{-1}$. The difference in luminosity to our sample is to be expected, as the JG19 sample was drawn from the shallower, but much wider, FIRST survey, and so represents relatively rare, higher luminosity objects. Using the higher sensitivity of LOFAR we are able to reveal for the first time the larger population of small, low-luminosity sources.

\subsubsection{Radio Morphology}

We used the LoMorph code of \citet{Mingo2019RevisitingLoTSS} to provide a systematic classification of our sources as either FRI or FRII, using the traditional definition of whether the peak flux is closer to the centre or to the edge of the source respectively. We find 67 FRI-type sources (65 and 16 within the AS and VS respectively) and 8 FRII-like sources (8 and 1 respectively). The remaining objects are either too small and faint to be classified automatically by LoMorph or have a mixed morphology (notably this includes ILTJ112543, discussed in Appendix~\ref{sec:AppendixPhotoSpecAnalysis}, which we visually classify as an FRII). Unfortunately, as Mingo et al. note, LoMorph is less reliable when applied to small, FRI-like objects. As a result we visually checked these sources, finding that a small number of GSJ have been misclassified (for example we classify ILTJ121847 as an FRII whilst LoMorph classifies it as an FRI), though overall we find qualitatively similar results, with the majority of FRI sources being correctly classified.

Though they acknowledge multiple exceptions, \citet{Best2009RadioSDSS} found that for luminosities above $10^{25}\text{ W Hz}^{-1}$ (at $1.4\text{ GHz}$), equivalent to $\sim10^{26}\text{ W Hz}^{-1}$ at $150\text{ MHz}$, the majority of sources are classed as FRII. However, all of our FRII-like sources have luminosities below this limit, which is consistent with the recent samples of lower-luminisoty FRIIs found by \citet{Mingo2019RevisitingLoTSS} and \citet{Capetti2017FRIIGalaxies}.

\subsubsection{Spectral Indices}
\label{sec:Results-Spectral}

In this section we consider the integrated spectral indices for our sample and compare them with the wider populations of compact and extended radio galaxies. Relationships between integrated spectral properties, radio power and size are expected on theoretical grounds, and have also been seen in large samples \citep[e.g.][]{deGasperin2018ASky,Tisanic2020TheNuclei}, although the effect of flux limits must be accounted for \citep[e.g.][]{Sabater2019TheOn}. Core-dominated and compact sources typically have flat integrated spectra at low frequencies, due to the effects of synchrotron self absorption and free-free absorption \citep[e.g.][]{ODea2020CompactSources}. For extended, optically thin sources, regions of flatter spectral index are associated with locations of recent particle acceleration \citep[e.g.][]{Heavens1987ParticleSpectrum} as are found at the base of FRI jets \citep[e.g.][]{Laing2013TheAcceleration}, whereas steep spectrum emission is typically associated with older plasma (although in some cases a comparatively steep injection spectrum is possible).

As already noted in Section~\ref{sec:size-selec}, our sources are all resolved at 150 MHz and, applying the criteria of \citet{Kellermann1981CompactSources}, they are all over 4 orders of magnitude too large to be optically thick so that any self absorbed component will be insignificant. Whilst our sources are also far larger than the intervening structures typically assumed to cause free-free absorption \citep[][]{ODea2020CompactSources}, simulations show that at low frequencies free-free absorption does become more prominent, especially in the core regions \citet{Bicknell2018RelativisticGalaxies}. Therefore, whilst the main cause of any change in the spectral slope of our sources is plasma age, some contribution from free-free absorption is possible.

To calculate the spectral index for each GSJ we used the $1.4\text{ GHz}$ fluxes from the NVSS \citep[][]{Condon1998TheSurvey} and FIRST \citep[][]{Becker1995TheCentimeters,White1997ASurvey} surveys. We cross-matched our sample of GSJ using a 20 arcsec search radius against NVSS and 5 arcsec search radius against the FIRST catalogue, before visually inspecting the matches to ensure they were referencing the same source. Multi-component objects, such as FRII-like sources, can have multiple catalogue entries and so we also manually checked the higher-resolution FIRST catalogue against all of our GSJ and if multiple components were found we used the cumulative flux.

FIRST has a limiting sensitivity of $1\text{ mJy}$, and so is more sensitive than NVSS at $2.5\text{ mJy}$. However, the largest angular size to which NVSS is sensitive is greater than that of FIRST. As a result, for those GSJ with a measured size of less than 30 arcsec across we use the FIRST values to calculate the spectral index. At our 30 arcsec limit, estimates show FIRST recovers $77$ per cent of the flux, though this rapidly increases for smaller sources \citep[see][for details]{Becker1995TheCentimeters}. For larger objects we use the measured NVSS fluxes. We calculate limits for objects undetected in either NVSS or FIRST.

\begin{figure}
    \centering
    \includegraphics[width=0.4\textwidth]{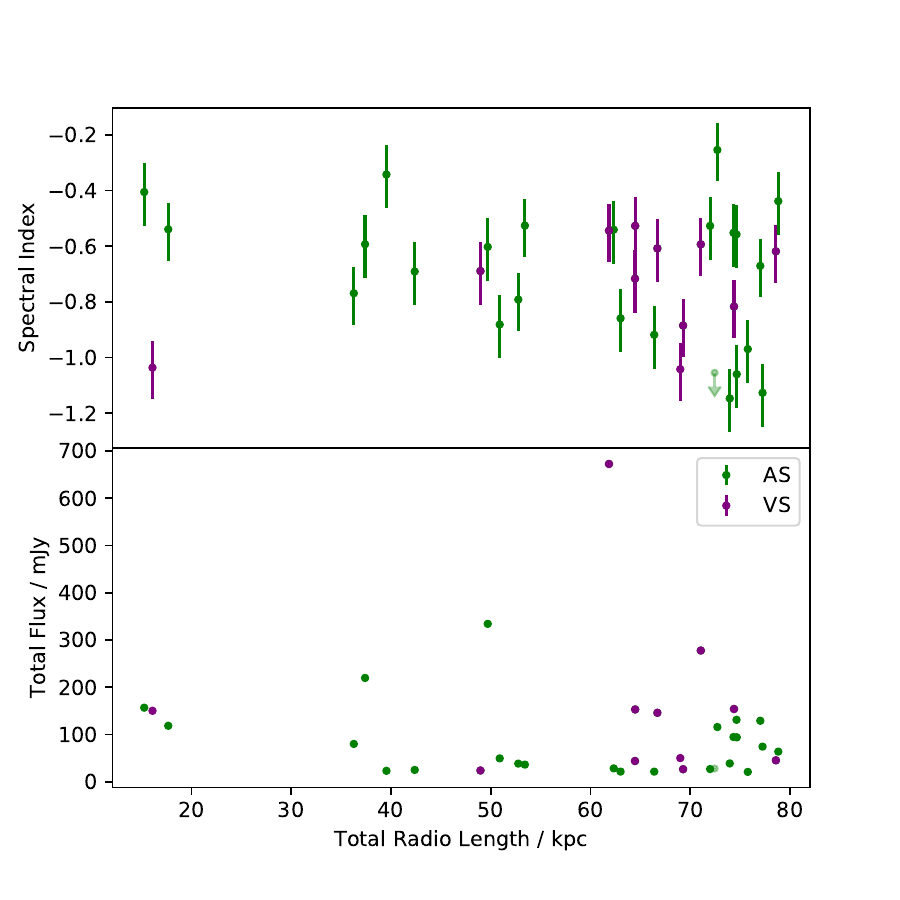}
    \caption{The top panel shows spectral index vs total radio length and the bottom panel shows 150 MHz total flux density vs total radio length. Only sources with a total flux greater than 20 mJy are shown. Though typically very small, errors in the total radio length have been omitted for clarity. For the faint source we could only derive an upper limit for the spectral index. Errors on the total flux are too small to be seen.}
    \label{fig:SpecIdx}
\end{figure}

For the 75 objects with 1.4 GHz detections, we find an average spectral index of $-0.60\pm0.12$. However, if we exclude those sources with fluxes below $20$ mJy, where the selection effects in the LoTSS sample are most prevalent \citep[see][for details]{Sabater2019TheOn}, the average becomes $-0.70\pm0.12$ (see top panel of Figure~\ref{fig:SpecIdx}). This is consistent with the values seen in more powerful radio-galaxy populations (e,g, the 3CRR sample of \citealt[][]{Laing1983BrightGalaxies}, with a typical spectral index of $\alpha = -0.76$), as well as the value of $-0.6 \pm 0.1$ found by \citet{Heesen2014ThePopulation} when studying a different GSJ, NGC 3801, and the value of $-0.63$ found by \citet{Sabater2019TheOn} when analysing the wider LoTSS AGN sample. Our spectral indices are consistent with the range of values found by JG19, though our average is slightly higher than the peak value they find of -0.5, possibly due to the hotspots in their small, luminous FRII sources being more dominant.

Resolved sources larger than a few hundred parsec typically have spectral indices steeper than $-0.5$. Therefore, using FIRST images, we visually examined the 23 sources with a spectral index flatter than $-0.5$. We find that they are all unresolved at the 1.4 GHz frequency of FIRST, so that for this subset of our sample the spectral index is dominated by the core properties and does not provide any information about source age or particle acceleration history. For the remaining sources, the integrated spectral index is providing a crude measure of age for the extended GSJ sources, with the comparatively steep spectrum sources likely to be older than those with flatter spectra, although we cannot rule out a contribution from free-free absorption for the smallest sources.

For the sources above 20 mJy, we find that those sources with steep spectral indices are more likely to have a large physical size within our sample range, with only one source with $\alpha<-0.9$ having a physical size less than $60\text{ kpc}.$ We test if the size-spectral index relation is caused by a relationship between flux and size (bottom panel of Figure~\ref{fig:SpecIdx}). However, we find no relationship between the flux and size showing that the spectral index-size relation present in our sample has a physical origin \citep[e.g.][]{Ker2012NewSamples}.

As a final comparison, we compare with the population of CSS sources -- these are physically small with a turnover in the spectral frequency, above which they have a steep spectral index. For example, \citet{ODea1998TheSources} defines a CSS as being smaller than 15 kpc with $\alpha < -0.5$. According to this criteria none of our GSJ are also classed as CSS sources, making the two categories distinct. This is likely due to a combination of the 12 arcsec cut and lower jet:galaxy ratio used during our sample selection (Section~\ref{sec:ratio-selec}) and so some overlap is anticipated once we are able to discover smaller GSJ in future, higher resolution surveys.

\subsection{Host Galaxies}
\label{sec:Results-Hosts}

We wish to compare the host properties of our GSJ with a larger sample of radio-galaxy hosts. When comparing photometric properties we use the same sample as Section~\ref{sec:Results-RadioProperties}, while for spectroscopic properties we use only the 170 GSJ with spectroscopic measurements, comparing these with the 2,544 such objects in the H19 sample, hereafter H19Spec. The comparisons discussed in this section are summarised in Table~\ref{tab:Properties} and Figure~\ref{fig:HostAndRadioProperties}.

\subsubsection{Host Morphology}
\label{sec:Results-Hosts-Morphology}

\begin{figure}
    \centering
    \includegraphics[width=0.4\textwidth]{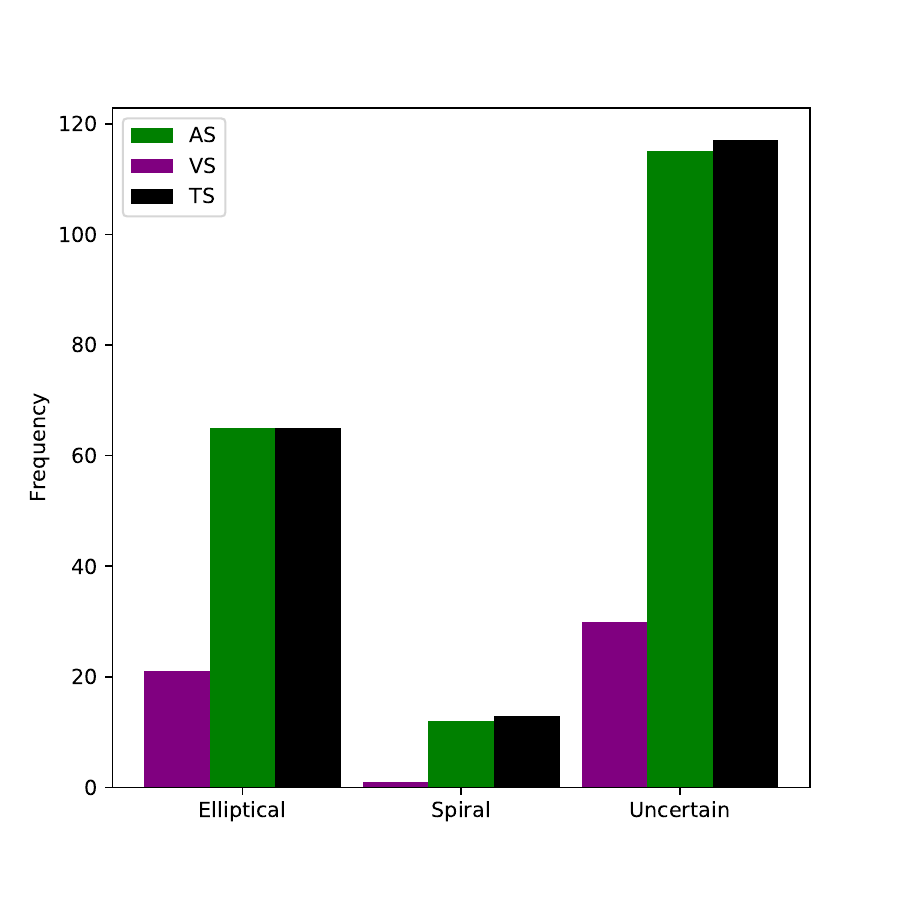}
    \caption{Host galaxy morphological classification from Galaxy Zoo.}
    \label{fig:GalZooHostType}
\end{figure}

To examine the host-galaxy morphologies for our sample we used the results of the Galaxy Zoo project \citep{Lintott2008GalaxySurvey}. We found that 13 of our GSJ were hosted by spirals: 12 are in the AS only and one, ILTJ121847, is in the VS only. As expected, the majority of hosts with a definite classification are ellipticals (see Figure~\ref{fig:GalZooHostType}). Applying our own visual classification to the indeterminate sources from Galaxy Zoo, we found an additional three spiral-hosted sources that appear in the AS only and one, ILTJ112543, that is in the VS only, resulting in 15 AS and 2 VS spiral-hosted sources.

Based on our visual inspection of the 15 AS spiral-hosted sources, eight appear to be star-forming galaxies of which three have strong radio cores. The radio emission from four of the spirals is Gaussian in shape whilst one has a continuous region of radio emission that appears to overlap with emission from background galaxies. Two sources have strong Gaussian radio emission on one side of the host only that may be from a background source, though no host galaxy can be seen. To test if these are contaminants we adopt the emission-line criteria of \citet{Kewley2006TheNuclei}, finding that none of these spiral-hosted sources are classified as strongly star-forming. In particular, the three sources with strong radio cores are classified as LINERs with an additional three being classified as Seyfert galaxies. Whilst this is not surprising as part of the test adopted by Kewley et al. (the [NII]/H$\alpha$ vs [OIII]/H$\beta$ BPT test) was also used by \citet{Hardcastle2019Radio-loudSources} when identifying radio-loud sources, the additional tests used by Kewley et al. show that even if there is some star-formation related emission present, these sources do have low levels of AGN activity and can be considered as GSJ.

The two spiral-hosted sources in the VS both exhibit strong FRII-like radio morphologies and are discussed in detail in Appendix~\ref{sec:AppendixAdditionalVSSSources}. These unusual objects belong to the class of so called spiral DRAGNs \citep[][]{Kaviraj2015RadioGalaxies}, with the luminosity of our sources being similar to the low-luminosity spiral DRAGN of \citet{Mulcahy2016TheDRAGN}.

Overall spirals comprise 4 per cent of the VS and 9 per cent of the TS. Whilst the results from the VS are consistent with those surveys conducted at higher frequencies which show that spiral hosts comprise less than 5 per cent of the total radio-loud population \citep[][]{Tadhunter2016RadioEvolution}, the fraction in the TS is marginally higher. Active nuclei in spirals are generally less powerful and so this increased percentage may be due to the majority of our sample being lower luminosity than that of Tadhunter. However, it may also be that spiral-hosted AGN are more easily detectable at low frequencies or that they are more likely to host GSJ.

We also looked for merger signatures, using the r-band images from the Pan-STARRS survey, but only found one source (ILTJ150245.73+533042.7) that shows any obvious signs of having undergone a recent merger. However, for many sources the optical image quality means it is impossible to rule this possibility out.

Finally, we also compared the concentration indices, C (where $C = R_{90}/R_{50}$), for our GSJ samples with the H19 sample. As shown in Table~\ref{tab:Properties}, the concentration values are consistent to within their errors, and consistent with expectations for elliptical/bulge-dominated galaxies which have concentration indices above 2.6  \citep[][]{Heckman2014TheUniverse}. The two spirals in the VS both have concentration indices about 2.2, typical of disk-dominated systems, but the 15 AS spirals have concentration indices ranging from 2.2 to 3.2 with a mean value of $2.6\pm0.3$. Therefore, whilst our spirals do have generally lower concentration indices than the elliptical hosts they are bordering on being considered bulge-dominated. This is also different to the results of JG19 who found no sources with concentration indices less than 2.86. This is likely to be due to the higher sensitivity of LOFAR detecting lower levels of emission from spiral/less bulge-dominated sources. 

\subsubsection{Colour and Magnitude}

We next considered the colour and brightness of the host galaxies. The u-r colour and absolute i-band magnitude, accounting for Galactic dust extinction and K correction \citep[][]{Chilingarian2010AnalyticalBands, Chilingarian2012AGalaxies}, are shown in Figure~\ref{fig:HostAndRadioProperties}, along with the normalised distribution for the H19 sample.

Following \citet{Kaviraj2015RadioGalaxies} who studied the colours of a selection of spiral-hosted AGN we define a `blue' galaxy as having a (u-r) colour less than $2.2$. Using the Agresti-Coull method \citep[][]{Agresti1998ApproximateProportions} to estimate the uncertainties we find that, as expected, our spiral hosts are generally bluer than the elliptical hosts with both VS spirals and $8^{+5}_{-4}$ per cent of the VS ellipticals classed as blue. The AS has $20^{+12}_{-8}$ per cent of the spirals and $12\pm3$ per cent of the ellipticals classed as blue. Overall, our GSJ have an average colour of about $2.7\pm0.7$, consistent with the average of $2.6\pm1.1$ for the H19 sample.

The absolute i-band magnitudes of our GSJ are typically between $-24$ and $-21$ with an average value of $-23.1\pm0.8$, identical to the H19 sample and consistent with the JG19 sample. Whilst our spirals are typically fainter with an average value of $-22.1\pm1.5$, they still fall within the same range of magnitudes as the elliptical hosts. The notable exception to this is the VS spiral, ILTJ112543 whose host is particularly faint with an absolute magnitude of approximately $-17.0$. The other VS spiral-hosted source also has an absolute magnitude of about $-20.4$, towards the lower end of our range. Both of these objects are discussed further in Appendix~\ref{sec:AppendixPhotoSpecAnalysis}.

\subsubsection{Stellar and Black Hole Masses}

Using SDSS stellar mass estimates, which were derived using the methods of \citet{Kauffmann2003StellarSurvey}, the TS has a mean stellar mass of $\sim10^{11.2\pm0.4} \text{ M}_{\odot}$, very close to the value of $\sim10^{11.3\pm0.3} \text{ M}_{\odot}$ found for the H19Spec sample. This is slightly lower than the characteristic value of $10^{11.5}\text{ M}_{\odot}$ identified by \citet{Heckman2014TheUniverse} at which the overall population of AGN switches from releasing energy primarily through radiation to jets, but is consistent with the range ($\sim 10^{11}$ to $10^{12}\text{ M}_{\odot}$) identified by Heckman and Best as being typical of radio-loud galaxies. The spiral-hosted sources do have a slightly lower average stellar mass of $\sim10^{10.7\pm0.8}\text{ M}_{\odot}$, but consistent within the large uncertainty for this smaller sub-sample. Figure~\ref{fig:HostAndRadioProperties} shows one significant outlier with an unusually low stellar mass of $\sim10^{8.6} \text{ M}_{\odot}$, allowing it to be classed as a dwarf galaxy \citep[][]{Yang2020AJ090613.77+561015.2}. This is the spiral-hosted source ILTJ112543. ILTJ121847, the other spiral-hosted VS source, is the next most massive source in our sample with a stellar mass of $\sim10^{9.8} \text{ M}_{\odot}$.

Black hole masses were estimated using the $\text{M-}\sigma$ relation of \citet{McConnell2013RevisitingProperties}. We excluded four objets from our spectroscopic sample due to having measured velocity dispersions below below the resolution limit of SDSS. We note that the velocity dispersion contamination effect due to overall galaxy rotation identified by \citet{Hasan2019GalaxiesZ} does not have an impact on our comparisons. The average estimated black hole mass for both the AS and VS are $10^{8.6\pm0.5}$ and $10^{8.8\pm0.4}\text{ M}_{\odot}$ respectively, with the H19Spec sample having an average of $10^{8.7\pm0.6}\text{ M}_{\odot}$. Our values are consistent with the average of $10^{8.5}\text{ M}_{\odot}$ found by JG19 and places these objects within the range of black hole masses identified by \citet{Heckman2014TheUniverse} of $10^8$ to $10^{9.5}\text{ M}_{\odot}$ as typical of radio-loud AGN. As expected, the AS spirals have lower black hole masses with an average of $10^{7.7\pm0.4}\text{ M}_{\odot}$, placing them on the boundary of what is typical of a radio-loud AGN. 

\subsubsection{Stellar Properties}

Spectroscopic sources within the SDSS database have estimates of the SFR derived using the methods of \citet{Brinchmann2004TheUniverse}. Although optical AGN activity can cause SFR estimates to be too high, this is unlikely to be significant for our GSJ sample, as most FRI-type radio galaxies, such as the majority of our sample, have very little nuclear line emission, making them optically similar to ordinary non-active galaxies. Further, Brinchmann et al., adapt their methods to account for those sources identified as hosting an AGN. The average SFR of the spectroscopic GSJ for the TS is $10^{-0.8\pm0.6}\text{ M}_{\odot}\text{ yr}^{-1}$, consistent with the  $10^{-0.6\pm0.6}\text{ M}_{\odot}\text{ yr}^{-1}$ of the H19Spec sample, and with the SFR of less than $10^{0.5}\text{ M}_{\odot}\text{ yr}^{-1}$ expected for radio-loud AGN not undergoing a starburst \citep[][]{Tadhunter2016RadioEvolution}.

We find no difference between SFRs for the AS and VS, but as expected, the AS spiral galaxies have higher star formation rates than our elliptical hosts ($10^{-0.2\pm1.0}\text{ M}_{\odot}\text{ yr}^{-1}$), albeit with large uncertainty. The major exception to this is the VS source ILTJ112543 which we classify as a spiral (though not identified as such by Galaxy Zoo) and which has a particularly low SFR just above $10^{-3.0}\text{ M}_{\odot}\text{ yr}^{-1}$.

\citet{Kauffmann2009FeastGalaxies} find that active star-forming galaxies have a $4000\text{ \AA}$ break strength less than 1.4 whilst passive galaxies have a break above 1.7. The average $4000\text{ \AA}$ break strength for the AS, VS and TS is around $1.9\pm0.2$ and is therefore fairly typical for an evolved population of hosts. As expected the AS spirals within our sample have a lower break strength of about $1.5\pm0.4$, but again consistent within uncertainties. Both VS spirals have lower $4000\text{ \AA}$ break values of $1.1$. This is different to the results of JG19 who only found one source with a $4000\text{ \AA}$ measurement less than 1.7. Again this difference with JG19 is likely to be due to differences between low and high luminosity host galaxies. 

Finally, we report a comparison of stellar surface mass density in Table~\ref{tab:Properties} and Figure~\ref{fig:HostAndRadioProperties}, again finding results in line with typical properties of elliptical galaxies \citep[e.g.][]{Heckman2014TheUniverse}. Consistent with their spiral nature, the two VS spirals both have smaller surface mass densities of $10^{7.5\pm0.0}$ and $10^{7.9\pm0.0}$ for ILTJ112543 and ILTJ121847.

\subsubsection{Summary}

\begin{figure*}
\centering
\begin{tabular}{cc}
\includegraphics[width=6.95cm,trim={0 0 0 25}, clip=true]{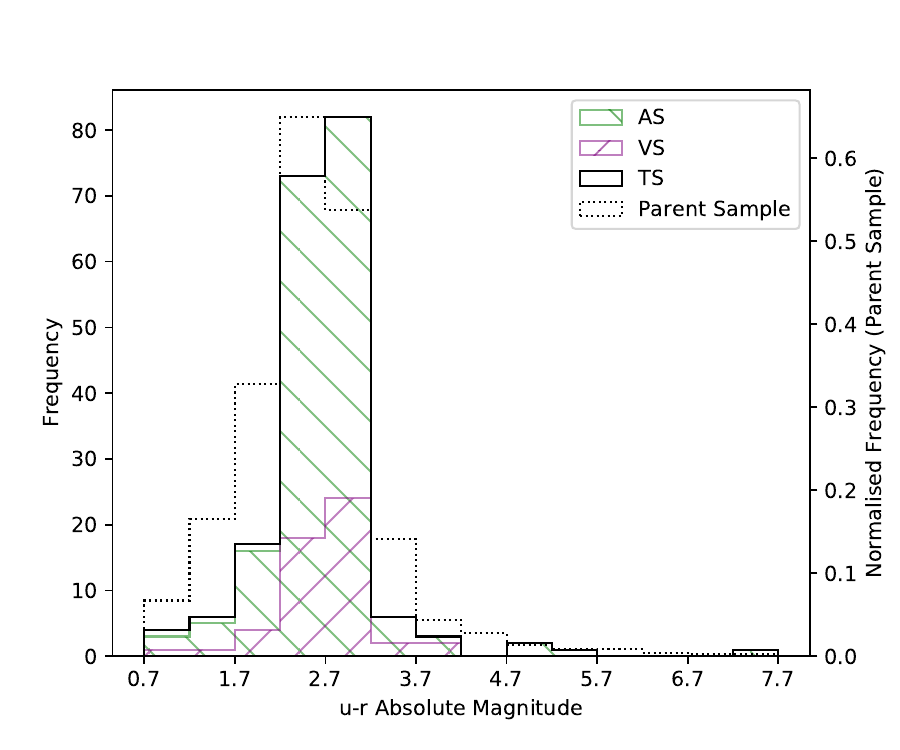} &
\includegraphics[width=6.95cm,trim={0 0 0 25}, clip=true]{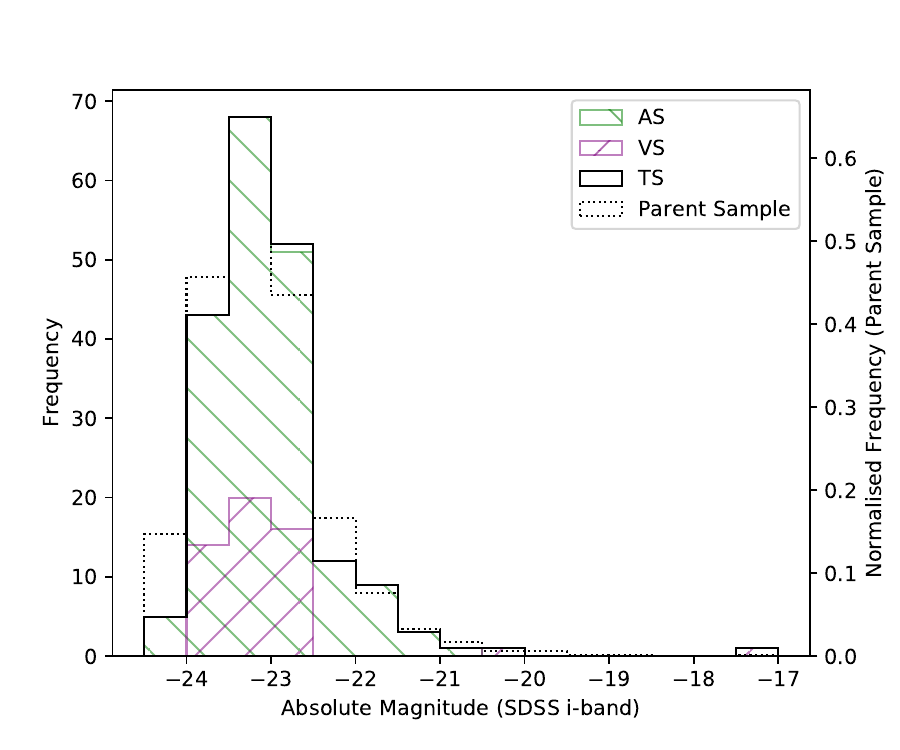} \\
\includegraphics[width=6.95cm,trim={0 0 0 25}, clip=true]{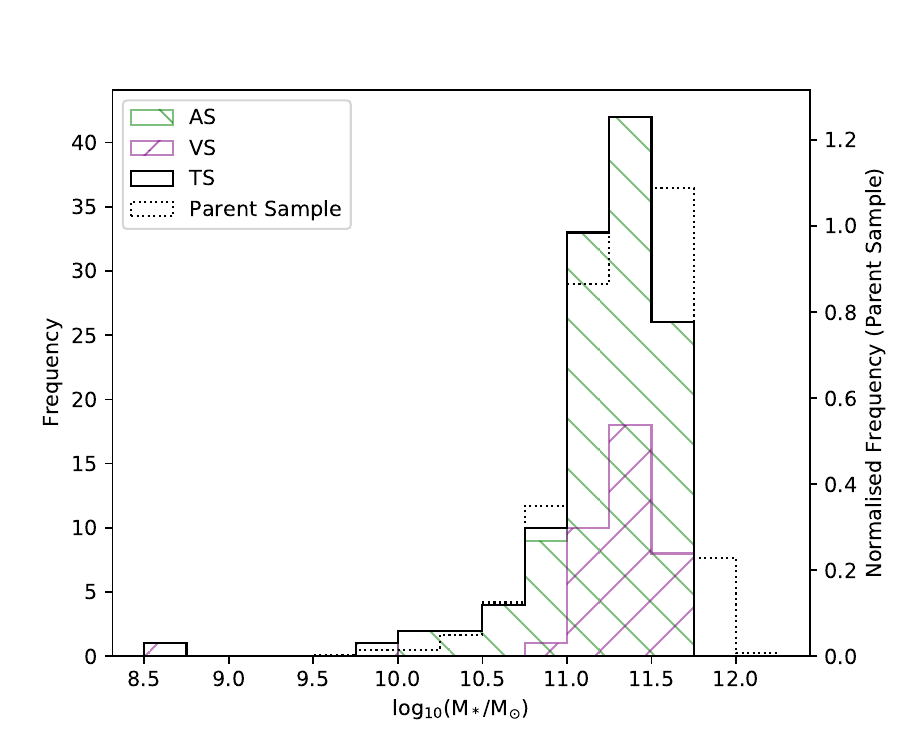} &
\includegraphics[width=6.95cm,trim={0 0 0 25}, clip=true]{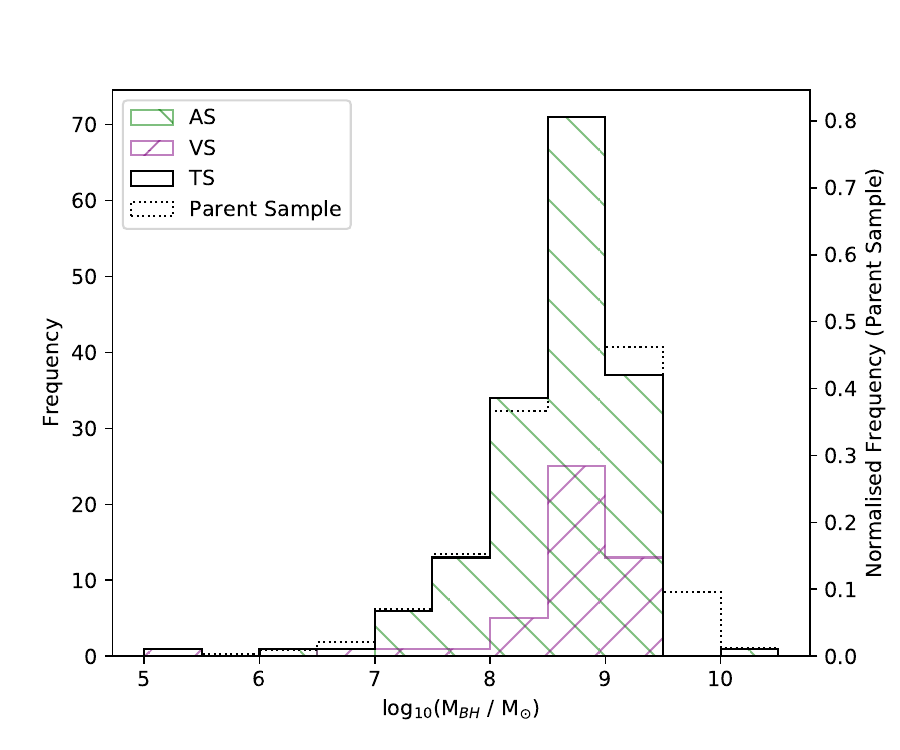} \\
\includegraphics[width=6.95cm,trim={0 0 0 25}, clip=true]{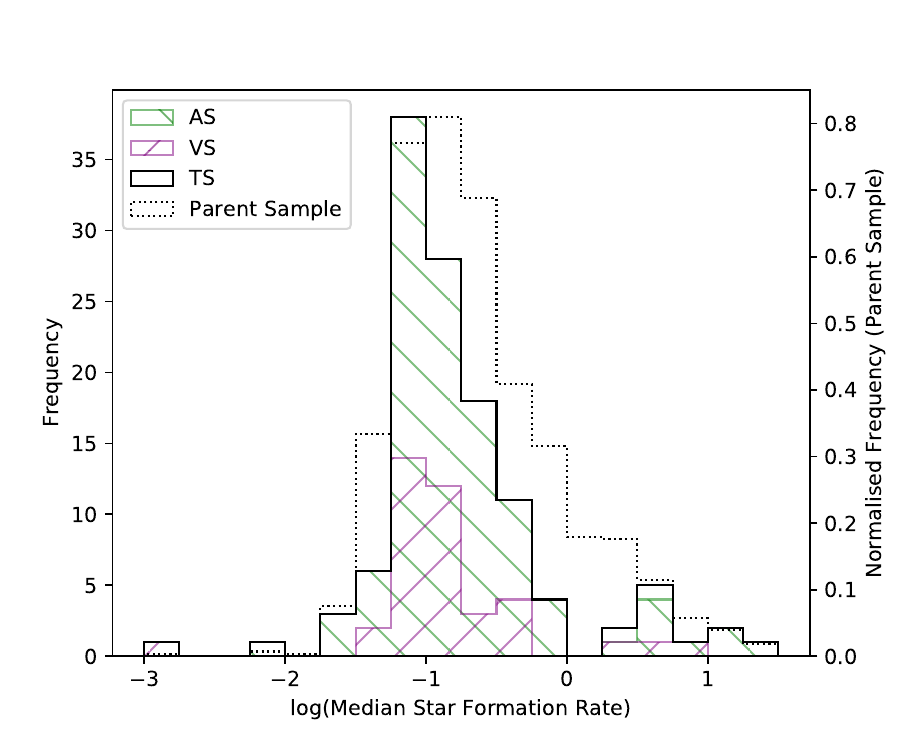} &
\includegraphics[width=6.95cm,trim={0 0 0 25}, clip=true]{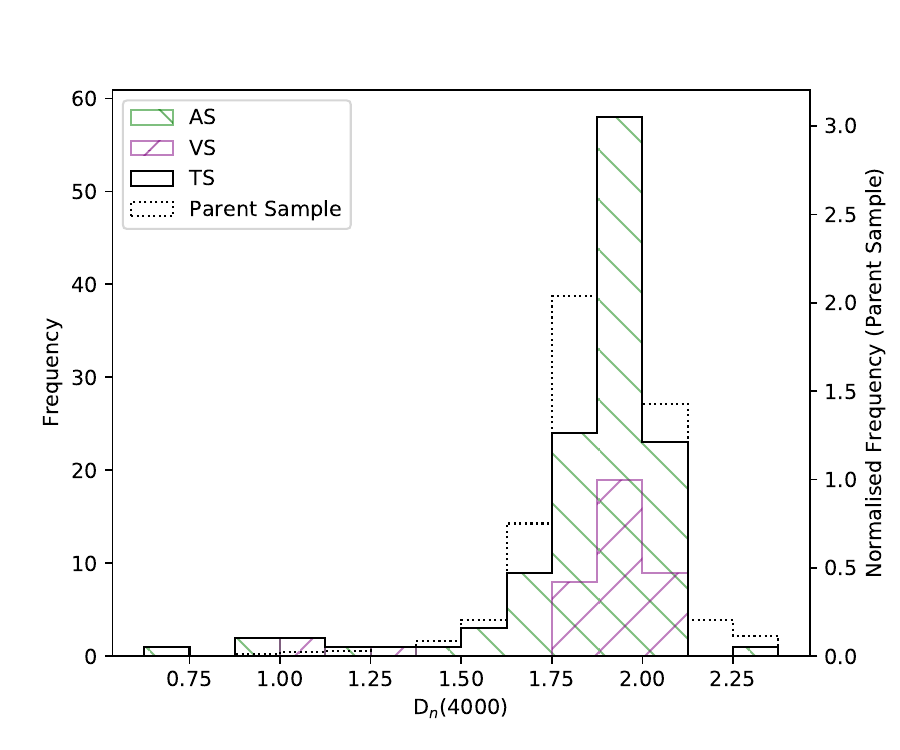} \\
\includegraphics[width=6.95cm,trim={0 0 0 25}, clip=true]{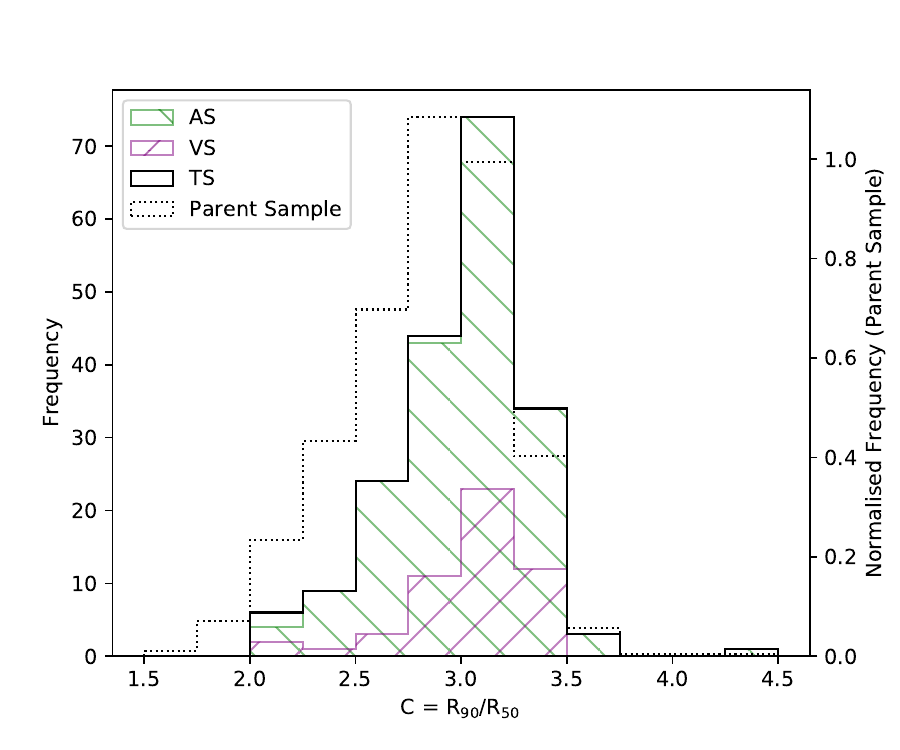} &
\includegraphics[width=6.95cm,trim={0 0 0 25}, clip=true]{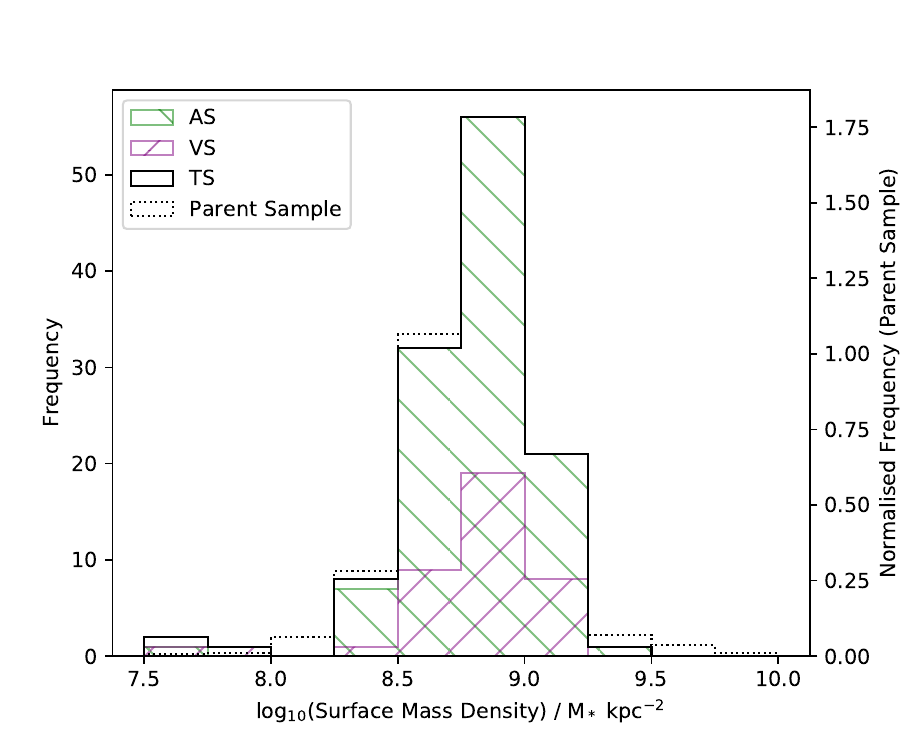}
\end{tabular}
\caption{Host galaxy properties for the TS, AS and VS. The black line in each diagram shows the H19/H19Spec parent sample, normalised to the TS. Top row: host galaxy colour (left) and absolute (i-band) magnitude (right). Second row: host stellar mass (left) and host black hole mass (right). Third row: host star formation rate (left) and 4000\AA{} break strength (right). Bottom row: concentration index, $C=R_{90}/R_{50}$ (left) and stellar surface mass density, $\mu_*=2\pi R_{50}^2$ (right).}
\label{fig:HostAndRadioProperties}
\end{figure*}

Figure~\ref{fig:HostAndRadioProperties} and Table~\ref{tab:Properties} demonstrate that the colour, absolute magnitude, stellar mass, black hole mass, SFR, 4000\AA{} break, concentration index and surface mass density for our GSJ samples all have similar mean values and distributions as the H19/H19Spec parent sample. GSJ are therefore hosted by galaxies that are typical of the broader radio-galaxy population. The number of spiral hosted GSJ is sufficiently small that this result is true even if we exclude these sources.

Even though the number of spirals within our sample is relatively small they do form a distinct subset within our population. The AS spirals have properties more typical of the wider population of spiral galaxies with higher host magnitudes, relatively blue colours, lower stellar and black hole masses, lower 4000\AA{} break strengths, lower surface mass densities, higher star formation rates and lower concentration indices. Individually these differences are marginal compared to the H19/H19Spec sample, with mean values for the spirals having large scatter. This may suggest that the central jet-generating regions of spiral-hosted GSJ are also similar to those of larger elliptically-hosted radio-loud AGN. In contrast, the two VS spirals are notable exceptions with significantly different host properties, making them particularly interesting objects for follow-up observations.

\begin{table*}
    \centering
    \begin{tabular}{l|c|c|c|c|c}
        \hline
        & TS & AS & AS(Spiral) & VS & H19/H19Spec\\
        \hline
        Host Magnitude (i-band)&$-23.1\pm0.8$&$-23.1\pm0.6$&$-22.5\pm0.8$&$-23.0\pm1.0$&$-23.1\pm0.8$\\
        Host Colour (u-r)&$2.7\pm0.7$&$2.7\pm0.6$&$2.3\pm0.6$&$2.7\pm0.5$&$2.6\pm1.1$\\
        Concentration Index ($R_{90}/R_{50}$)&$3.0\pm0.3$&$3.0\pm0.3$&$2.6\pm0.3$&$3.1\pm0.3$&$2.8\pm0.4$\\
        Stellar Mass (log$_{10}$(M$_*$/M$_{\odot}$))&$11.2\pm0.4$&$11.3\pm0.3$&$10.9\pm0.4$&$11.2\pm0.5$&$11.3\pm0.3$\\
        BH Mass (log$_{10}$(M$_{\text{BH}}$/M$_{\odot}$))&$8.6\pm0.5$&$8.6\pm0.5$&$7.7\pm0.4$&$8.8\pm0.4$&$8.7\pm0.6$\\
        Median Star Formation Rate&$-0.8\pm0.6$&$-0.8\pm0.6$&$-0.2\pm1.0$&$-0.9\pm0.6$&$-0.6\pm0.6$\\
        4000\AA{} Break&$1.9\pm0.2$&$1.9\pm0.2$&$1.5\pm0.4$&$1.9\pm0.2$&$1.9\pm0.3$\\
        Surface Mass Density (M$_*\text{ kpc}^{-2}$)&$8.8\pm0.3$&$8.8\pm0.2$&$8.6\pm0.4$&$8.8\pm0.3$&$8.8\pm0.3$\\
        \hline
    \end{tabular}
    \caption{Comparison of the host properties of the TS, AS and VS with the \citet{Hardcastle2019Radio-loudSources} sample. Also shown is the subset of spiral galaxies from the AS.\bigbreak}
    
    \label{tab:Properties}
\end{table*}

\subsection{Environmental Richness}
\label{sec:Results-Environments}

To investigate the large-scale environments of the GSJ, we used the catalogue of \citet{Croston2019TheLoTSS} which cross-matches LoTSS AGN with the SDSS cluster catalogues of \citet{Wen2012AIII} and \citet{Rykoff2014RedMaPPer.CATALOG} to estimate cluster richness. Adopting a matching probability greater than 80 per cent, 17 of our GSJ have a match in the catalogue of Rykoff et al. and 38 have a match in the catalogue of Wen et al., with 13 having a match in both. We therefore report our results using matches from the Wen catalogue, though we note that our results are qualitatively the same when using either catalogue.

\begin{figure}
\centering
\includegraphics[width=0.4\textwidth, trim={20 0 50 50}]{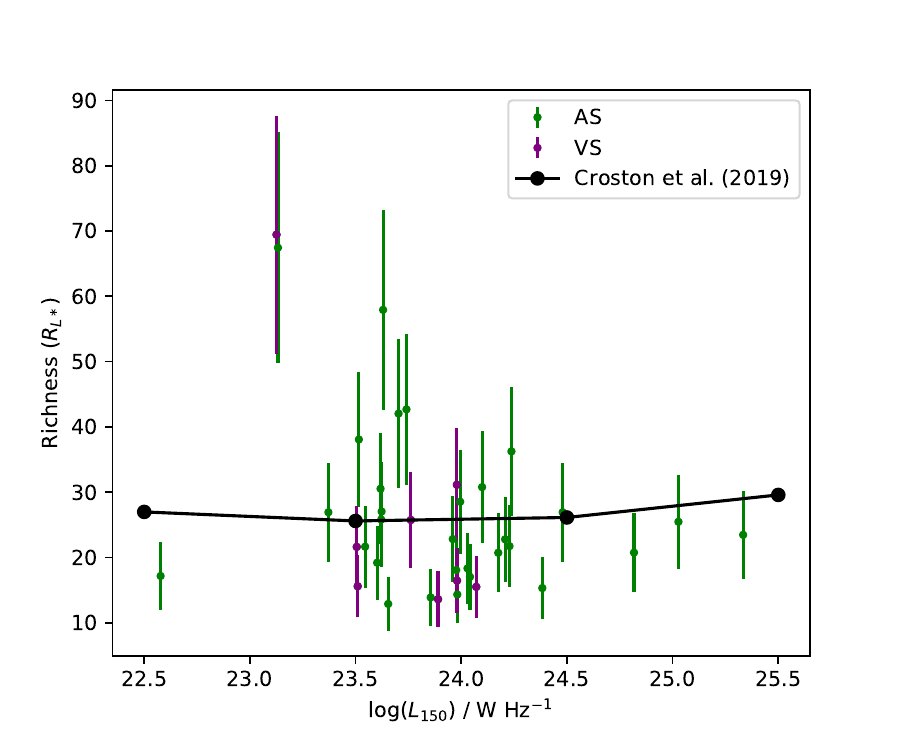}
\caption{The size of the clusters associated with our GSJ as given by \citet{Wen2012AIII}. The black lines show the average cluster size/richness as a function of radio luminosity for matched sources as found by \citet{Croston2019TheLoTSS}.}
\label{fig:GSJClusters}
\end{figure}

Those GSJ with a match are shown in Figure~\ref{fig:GSJClusters}, where we use the $R_{L*}$ richness indicator of Wen at al, which, they define as the approximate number of cluster galaxies within the $r_{200}$ radius. Those sources with a match are broadly consistent with the average relationship found by Croston et al. between cluster size and radio luminosity. The majority of GSJ with a cluster match are located near the catalogued cluster centre, however, those GSJ in larger groups tend to be found away from the cluster centre. Along with the lack of any secondary galaxies in the majority of our cutout images (see Appendix~\ref{sec:AppendixVSSSourceImages}) this indicates that our matched GSJ are observed predominantly in relatively poor/sparse environments.

The majority of our GSJ do not have a match. For these sources we note that, at a redshift of 0.42, the Wen at al. catalogue is > 95 per cent complete above $M_{200} > 10^{14}\text{ M}_{\odot}$ for clusters with a size $R_{L*}\geq12$ whilst at higher redshifts the cluster sizes are likely to be under-reported. Of our 157 unmatched sources, only five have a redshift greater than 0.42 meaning our unmatched sources are also located in poor/sparse environments. Our GSJ are therefore found in similar environments to the FR0 population, which are typically found in groups of less than 15 galaxies \citep[][]{Capetti2020Large-scaleGalaxies}.

\section{Prevalence of GSJ}
\label{sec:Results-Fraction}

It is well established that the likelihood of a galaxy hosting a radio-loud AGN of any given luminosity increases with both the host's stellar and black hole masses \citep[][]{Best2005ASurvey,Sabater2019TheOn}. In order to determine whether the same relationship is true of our GSJ we adopted the techniques of \citet{Sabater2019TheOn} and calculated the fraction of the total number of galaxies that are GSJ at the current time. We also investigate whether the fraction of radio galaxies that we have classified as GSJ is similarly dependent upon the host's stellar and black hole mass. Although the resolution of LOFAR means that as redshift increases we are unable to observe the smallest GSJ, we found that accounting for this does not qualitatively affect our results and so we make no adjustments to our sample size. The results presented in this Section represent a lower limit on the prevalence of GSJs within the radio-galaxy and overall nearby galaxy populations.

To compare our sample of GSJ with the wider population of galaxies and to allow a direct comparison with the work of Sabater et al. we took all those galaxies from within the SDSS Main Galaxy Sample that are located within the HETDEX footprint. The Main Galaxy Sample has an approximate upper redshift limit of 0.3, which, in order to produce an unbiased comparison we use as the upper limit within this section, resulting in a GSJ sample size of 165. We divided both samples into bins of stellar and black hole mass (derived as described in Section~\ref{sec:Results-Hosts}). For each bin we calculated the fraction of galaxies that are GSJ where both have a luminosity $\geq 10^{21}\text{ W Hz}^{-1}$. We then repeated the processes, increasing the luminosity limit by one dex each time up to a maximum of $10^{24}\text{ W Hz}^{-1}$.

\begin{figure*}
    \centering
    \begin{tabular}{cc}
        \includegraphics[width=0.4\textwidth]{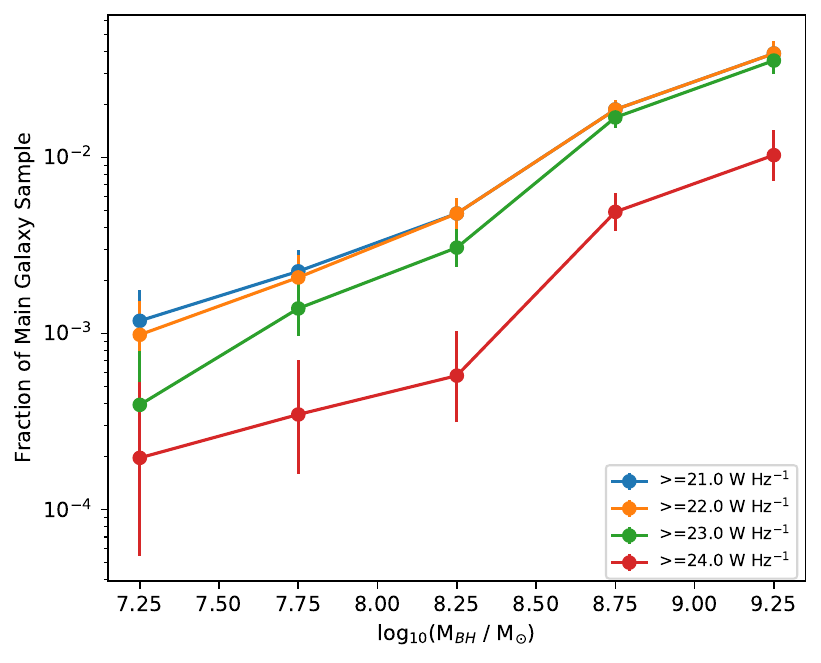}&
        \includegraphics[width=0.4\textwidth]{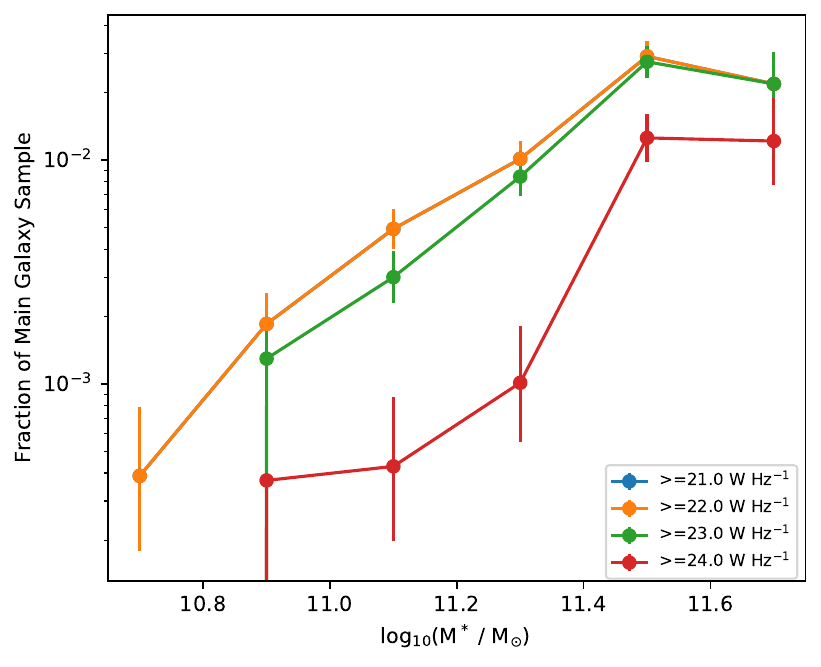} \\
        \includegraphics[width=0.4\textwidth]{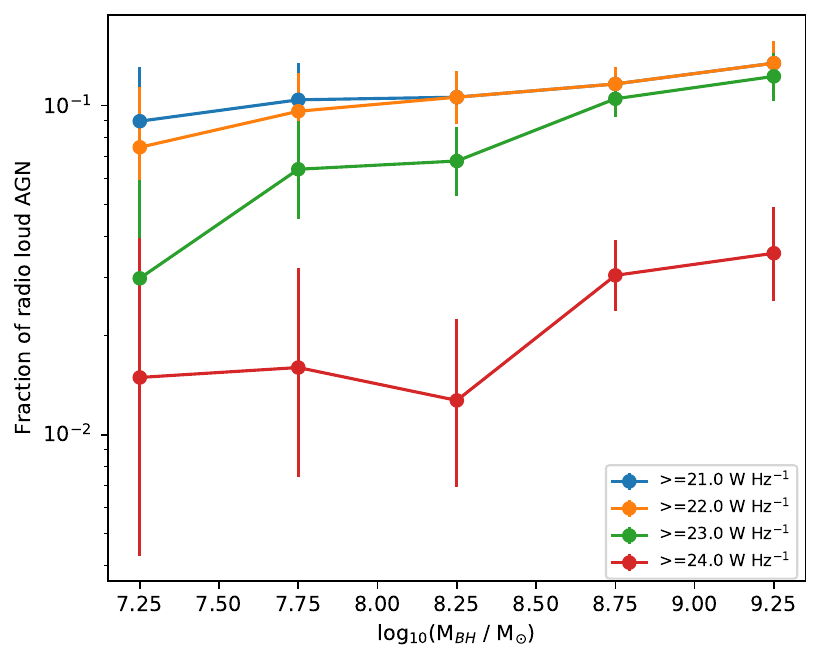} &
        \includegraphics[width=0.4\textwidth]{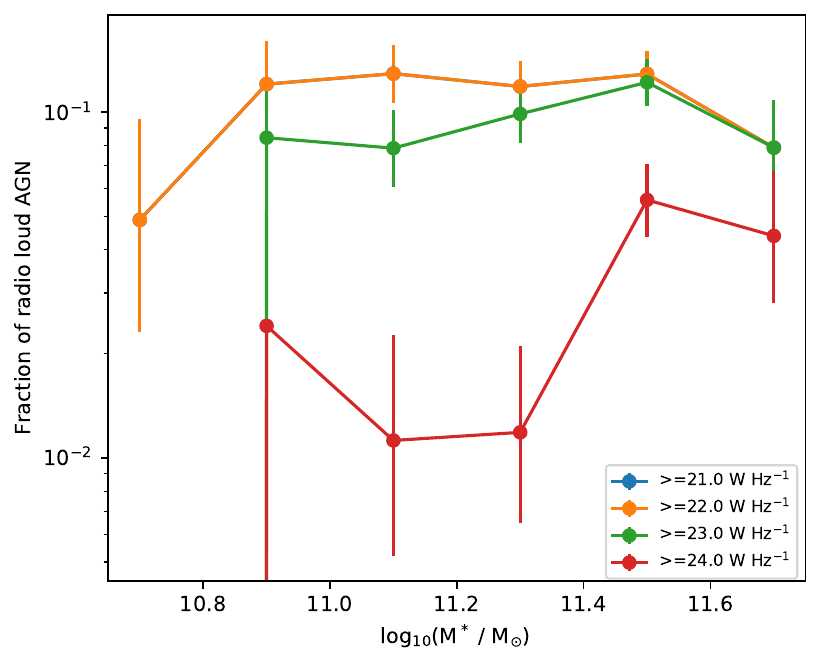} \\
    \end{tabular}
    \caption{Top Row: Fraction of Main Galaxy Sample galaxies that are GSJ with luminosity above the given limits as a function of black hole mass (left) and stellar mass (right). Bottom Row: Fraction of radio-loud galaxies that are GSJ with luminosity above the given limits as a function of black hole mass (left) and black hole mass (right). The percentages shown represent lower limits as no adjustment has been made for selection effects.}
    \label{fig:GSJFrac}
\end{figure*}

The results are shown in the top row of Figure~\ref{fig:GSJFrac}. The uncertainties shown were calculated using the Agresti-Coull binomial confidence interval \citep[][]{Agresti1998ApproximateProportions} with the confidence level set at $68$ per cent. For clarity, and because their numbers are too small to be statistically useful, we exclude those galaxies with stellar masses below $10^{10.6}\text{ M}_{\odot}$ from the plots (see Figure~\ref{fig:HostAndRadioProperties}).

When compared to the wider population of galaxies (upper panels) we find that the fraction of GSJ is directly related to both the black hole and stellar mass. This is entirely consistent with the conclusions of \citet{Sabater2019TheOn}. In contrast to their results, however, we note that, for all luminosities, the relation with stellar mass appears to flatten above $10^{11.5}\text{ M}_{\odot}$. We tested the significance of this flattening for the most inclusive luminosity group ($L_{150}>10^{21}\text{ W Hz}^{-1}$) by applying a linear fit to the points up to $10^{11.5}\text{ M}_{\odot}$. In the absence of a change of slope, we would predict the fraction with stellar masses of $10^{11.7}\text{ M}_{\odot}$ that are also GSJ to be $0.09\pm0.03$. Since the actual data point is $0.02\pm0.01$, the flattening is significant at more than two sigma. With the exception of the highest luminosity group ($L_{150}>10^{24}\text{ W Hz}^{-1}$), which has a smaller number of sources and is therefore less reliable, the other luminosity groups also show a flattening at more than the two sigma level. We consider the origin of this flattening, not seen for the larger sample of \citet{Sabater2019TheOn} in Section~\ref{sec:Discussion}. 

As well as investigating the prevalence of GSJ in the Main Galaxy Sample, we carried out a similar comparison using a parent sample of radio-loud AGN, allowing us to investigate the fraction of radio-loud AGN that we have classified as GSJ. For consistency we use the parent sample of Sabater et al. rather than that of \citet{Hardcastle2019Radio-loudSources}. The sample of Sabater et al. is limited to a redshift of 0.3, allowing a direct comparison with the above results. The results are shown in the bottom two plots of Figure~\ref{fig:GSJFrac}. We find that the fraction of radio-loud AGN that are GSJ is independent of black hole mass, remaining roughly constant at $\sim 10$ per cent for GSJ with luminosities above $10^{21}\text{ W Hz}^{-1}$. The fraction of GSJ also remains broadly constant with respect to stellar mass for all except the most luminous group ($L_{150}>10^{21}\text{ W Hz}^{-1}$), which is less reliable due to the smaller number of objects. Whilst there is a slight drop at a stellar mass of $10^{11.7}\text{ M}_{\odot}$, this is only significant at the 1.5 sigma level. Further studies are needed to see if the fraction of radio-loud galaxies that are GSJ does decrease at the highest stellar and black hole masses.

\section{Energetics}
\label{sec:Results-Energetics}

To get a first-order estimate of the impact our GSJ could be having upon their hosts we compare the internal energy within the radio lobes with the energy within the host's hot ISM. As mentioned in Section~\ref{sec:TheData} the ellipse sizes from the LoTSS catalogue typically under-represent the true size of our sources. Therefore, rather than using these sizes to calculate the radio energy we instead assume the radio emission comes from a cylindrical region of typical aspect ratio. We used the VS to determine a typical aspect ratio, finding that the source diameter is typically $\sim 0.55\pm0.12$ times the length. Assuming this ratio applies across our entire sample, we used the source length and estimated diameter to estimate the radio-emitting volume, and using a Python version of the SYNCH code of \citet{Hardcastle1998Magnetic111}\footnote{https://github.com/mhardcastle/pysynch} we derived the minimum energy density and hence minimum total energy for each source.

\begin{figure}
    \centering
    \includegraphics[width=0.4\textwidth, trim={10 10 20 40},clip]{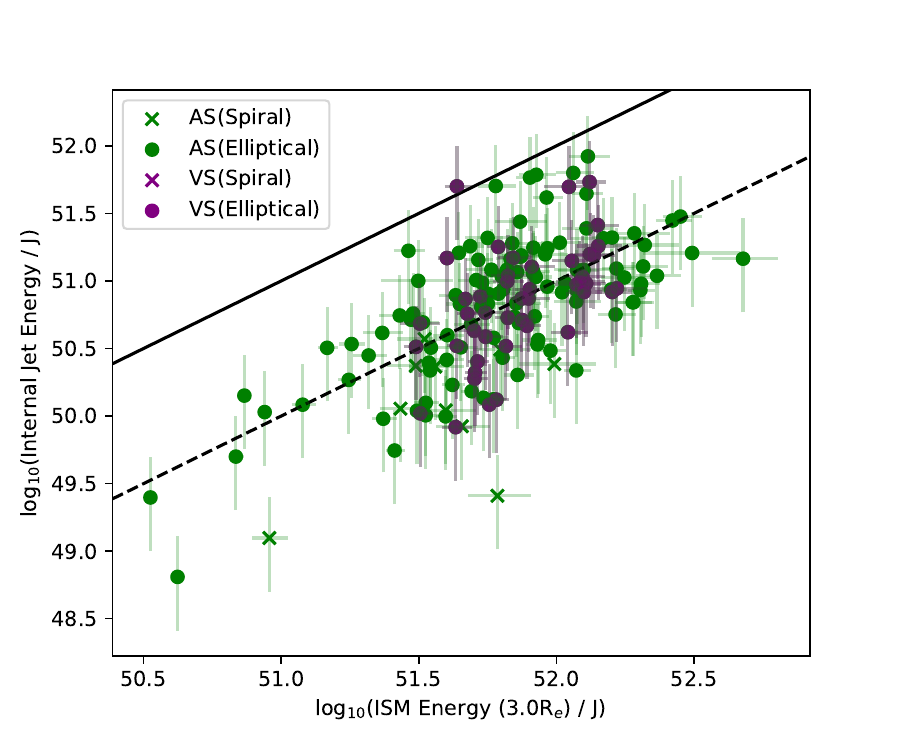}
    \caption{Comparison of the total internal radio energy against the energy of the ISM within three effective radii of the host. The solid diagonal black line illustrates where the internal radio energy equals the ISM energy within $3\text{R}_e$, whilst the dashed line shows where the jet energy is one tenth of the $3\text{R}_e$ ISM energy.}
    \label{fig:EnergyComparison}
\end{figure}

For powerful FRII sources it has been demonstrated that energy estimates within a small factor of equipartition give lobe pressures consistent with those required to inflate the lobes and achieve pressure balance with the environment \citep[e.g.][]{Ineson2017AGalaxies}. However for FRI-like sources the minimum energy estimates are often insufficient to inflate the observed cavities. This discrepancy is commonly attributed to the entrainment of protons. Based on the results of \citet{Croston2008AnGalaxies} and \citet{Croston2018ParticleEqual} we increased the minimum energy estimates for our FRI sources by a factor of 10 to better approximate the true energy within the lobes. We find that for the majority of our sources the internal energy of our GSJ ranges from approximately $10^{49.5}\text{ J}$ to $10^{51.5}\text{ J}$ (see Figure~\ref{fig:EnergyComparison}), indicating that even the least powerful GSJ contain almost a million times more energy than an average supernova of $10^{44}\text{ J}$.

However these minimum energy estimates represent only the internal energy of the lobes. The total energy available must be higher, to account for the work done in displacing the ISM as well as the existence of any shocks. For a relativistic gas undergoing adiabatic expansion, the enthalpy is $4pV$ \citep[e.g.][]{Birzan2004AGalaxies, Heckman2014TheUniverse}, although if shocks are present this figure could be higher \citep[e.g. in][the galaxy-scale jet source NGC 3801 was found to have an energy of up to $6pV$]{Croston2007Shock3801}. Consequently, an amount of energy equal to at least a third of the observed internal energy has already been transferred to the ISM. As discussed in, for example, \citet{Hardcastle2013NumericalEnvironments}, if shocks are present this figure could be significantly higher (although in that case our internal pressure estimate will be significantly higher than the external pressure relevant for $pV$ estimates).

To find the energy within the hot ISM we estimated the total gravitational mass of each host, using the velocity dispersion relations of \citet{Bandara2009AGALAXY} for the elliptical-hosted GSJ (Equation 7 in their paper) and \citet{Davis2019ARelations} for the spirals. We use these relations because our GSJ are predominantly in sparse environments (see Section~\ref{sec:Results-Environments}), and both authors use a selection of individual galaxies to find a direct relation between a galaxy's total mass and velocity dispersion. We limit ourselves to those sources flagged by SDSS as having reliable spectroscopic measurements. Our GSJ are, on average, slightly larger than three times the effective radius. Therefore, assuming a Navarro, Frenk and White (NFW) profile \citep[][]{Navarro1996TheHalos} with a concentration index of 6, we estimated the total mass within three effective radii of the host which is approximately the scale of influence of our GSJ and is the region we are most interested in. 

Assuming a fixed gas mass, we used the median value of $0.047\pm0.009$ found by \citet[][]{Dai2010ONGALAXIES} using a large sample of groups/clusters (richness class 2 in their terminology) to find the gas mass within each of our hosts. We further assume that the gas mass ratio derived by Dai et al. for groups/clusters can be applied to individual galaxies. However, we note that the gas mass fractions found by Dai et al. are consistent with \citet{Trinchieri2012HotiXMM-Newton/i} who, using X-ray data, found a gas mass fraction of 5 per cent for two galaxies with similar masses and environments to ours. We note that this is a conservative assumption, as the impact of stellar and AGN feedback process that may expel gas will be stronger on galaxy compared to group scales, so that gas fractions may be lower (as is the case for the Milky Way, which has an estimated hot gas mass fraction of around 2--3 percent \citep[][]{Bland-Hawthorn2016TheProperties}), but are unlikely to be substantially higher. Our ISM energies are therefore unlikely to be systematically underestimated. We then assumed an average particle mass of $0.62\text{ M}_H$ \citep[][]{Goulding2016THEUNIVERSE} to get the total number of particles within the ISM. Assuming a gas temperature of 0.7 keV for the spirals \citep[][]{Li2017TheGalaxies} and 0.5 keV for the ellipticals \citep[][]{Goulding2016THEUNIVERSE} we were then able to derive estimates of the total ISM energy within $3\text{R}_e$ of the host.

Figure~\ref{fig:EnergyComparison} shows the comparison of internal radio and ISM energies. 
Whilst the majority of our sources are clustered towards the top right of our plot there are a few sources with lower internal and ISM energies located towards the bottom left. These are typically lower-luminosity sources and may represent a wider population of extremely low luminosity GSJ that we are currently unable to observe.

Figure~\ref{fig:EnergyComparison} also shows that whilst the majority of our GSJ have internal radio lobe energies less than the ISM energy within $3\text{R}_e$ of the host, there are a few with similar internal radio lobe and ISM energy. There are also nearly 50 per cent with an internal radio lobe energy that is within an order of magnitude of the ISM energy. This suggests that even ignoring any shocks, GSJ are capable of significantly affecting the evolution of the ISM within their own host galaxy.

Previous studies of a small number of GSJ show that they can transfer large amounts of energy from the lobes directly into the ISM \citep[][]{Croston2007Shock3801,Croston2009High-energyA,Mingo2011MarkarianSeyfert,Hota2012NGCFeedback}. Whilst we do not have any evidence of any direct coupling between the lobes and host ISM for our GSJ, if the lobes are in pressure balance with their environments, then between 1/3 and 2/3 of the internal energy (depending on whether the lobes are dominated by relativistic or thermal material) must already have been transferred to the environment. We cannot determine at what radius this occurred, but their small physical size means that GSJ must have already had a significant impact on their hosts. Unfortunately the future impact of GSJ is harder to determine as we do not know at what radius, or over what timescales the lobes will release their current energy. Previous studies have shown that GSJ can generate shocks \citep[][]{Croston2007Shock3801,Croston2009High-energyA,Mingo2011MarkarianSeyfert,Mingo2012ShocksGalaxy}. Although currently unknown, if shocks are present within our sample this would mean the current impact of GSJ on their hosts is likely to be significantly higher than our estimates. Future X-ray studies of GSJ are essential to detect any shocks present and obtain direct evidence for the coupling of jet energy on galactic scales.

\section{Discussion}
\label{sec:Discussion}

We have developed a method for finding GSJ that identified 454 candidates from among the initial 318,520 sources in LoTSS DR1. From this sample we used the AGN/star formation separation criteria of \citet{Hardcastle2019Radio-loudSources} to automatically select 192 GSJ. Separately, we also visually inspected the 454 sources identifying 52 GSJ. These samples comprise the AS and VS respectively. Combining the unique sources from each gives the TS of 195 GSJ. Below we discuss the implications of our results.

\subsection{Comparison with other work}

The large number of sources found by modern surveys, such as LoTSS, mean that methods such as those outlined in Section~\ref{sec:TheData} will be vital if we are to find more GSJ in future and ongoing surveys. All our GSJ are resolved at 150 MHz with radio emission extending beyond the central confines of the host. Our sample have two-sided jet lengths between $10\text{ kpc}$ and our $80\text{ kpc}$ upper limit. These sources are therefore physically bigger than the FR0 class of objects \citep[][]{Baldi2018FR0Galaxies} and the jetted radio cores seen in the LeMMINGS survey \citep[][]{Baldi2018LeMMINGs.Cores}. Future, higher resolution, releases of the LoTSS survey will allow identification of smaller GSJ amongst the population of currently unresolved sources. A larger sample of GSJ will not only allow better comparisons with these other populations of small sources, it will also provide the numbers necessary to investigate any evolution in the properties and types of GSJ hosts with redshift.

Within our sample we find that a significant minority are producing FRII-like structures typically found in more luminous radio galaxies. However, all our FRII sources have luminosities below the limit of $10^{26}\text{ W Hz}^{-1}$ (at 150 MHz) identified by \citet{Best2009RadioSDSS} as the point above which FRII sources are typically observed. Our sources are therefore consistent with the results of \citet{Mingo2019RevisitingLoTSS} and \citet{Capetti2017FRIIGalaxies}, both of whom found populations of FRII sources with luminosities below this limit. Mingo et al. found that the hosts of their low-luminosity FRII sample have lower absolute K-band magnitudes and therefore have lower masses than the hosts of both more luminous FRII sources and the FRI sources of equivalent size and radio luminosity. Whilst the stellar masses of our FRII sources are all below our sample average, our sample is too small to confirm this finding of Mingo et al.

In their recent paper \citet{Jimenez-Gallardo2019COMP2CAT:Universe} found a population of 43 FRII sources no larger than $30\text{ kpc}$ in size making them comparable to our GSJ. All their sources have luminosity below the traditional FRI/FRII divide. Their sample was identified from FIRST images of 3,357 sources taken from the catalogue of \citet{Best2012OnProperties} with redshifts less than 0.15. Their sample is however comprised entirely of FRII sources and whilst they do find more FRIIs this can be attributed to the larger sky area over which they searched, plus the possibility of contaminants within their sample. Their sources are also more luminous than ours and so they provide a complementary view of the luminous end of the GSJ population. 

\subsection{Spiral-hosted GSJ}

We find that between 4 per cent of the VS and 9 per cent of the AS have spiral hosts, though it is possible that some of the AS spirals are contaminants. Whilst all our spiral-hosted GSJ have properties fairly typical of spirals in general,  being younger and bluer in colour with smaller masses than the elliptical hosts, this is particularly true for our two VS spirals. This is, however, different from the population of spiral-hosted AGN found by \citet{Kaviraj2015RadioGalaxies} where $\sim90$ per cent had a red colour akin to elliptical-hosted AGN, albeit at higher radio luminosities than our sample. Kaviraj et al. also found a high incidence of mergers amongst their population which we do not observe in our sample, although image resolution makes it impossible to rule this possibility out. 

Our findings therefore support the prediction by \citet{Mulcahy2016TheDRAGN} of a population of low-luminosity, spiral-hosted radio-loud AGN. The lack of any obvious mergers within our sample also suggests that secular processes may be responsible for triggering low-luminosity AGN activity \citep[as suggested by authors such as][]{Man2019TheSDSS, Tadhunter2016RadioEvolution} and that in order for these objects to attain the higher luminosities seen in the samples of Kaviraj et al. an event, such as a merger, is necessary to increase the flow of fuel to the AGN. To study these unusual objects properly, it is important to identify other spiral-hosted radio galaxies in future wider area surveys. However, our results show that care must be taken when looking for these objects as traditional selection techniques may miss them.

\subsection{GSJ prevalence and relation to the wider radio-galaxy population}

There is a well-established link between both stellar and black hole mass and the fraction of galaxies hosting a radio-loud AGN. This trend has recently been confirmed within the LoTSS DR1 sample \citep[][]{Sabater2019TheOn}. As discussed in Section~\ref{sec:Results-Fraction}, our GSJ follow the trend for black hole mass at all masses and they follow the trend for stellar masses up to about $10^{11.5} \text{ M}_{\odot}$, above which there is some evidence that the fraction of GSJ starts to flatten contrary to what is seen in the radio-loud AGN population, though further studies are needed to confirm this trend. 

Whilst the observed flattening may be caused by a selection effect we have not accounted for, if genuine, it may be related to our findings in Section~\ref{sec:Results-Environments} that GSJ inhabit relatively poor environments. Using the LoTSS sample \citet{Sabater2019TheOn} established that at the frequencies observed by LOFAR the largest galaxies are always active. If these jets are always turned on then the constant injection of energy could mean that at the observed frequencies and resolution of LOFAR the radio emission from the two jets remains larger than the $80\text{ kpc}$ limit we defined for our GSJ. As a result it might be expected that there would be fewer GSJ found in dense environments and at these high stellar masses. 

The constant fraction of the radio-loud AGN population that can be classed as GSJ also shows that for all radio-loud AGN exhibiting a duty cycle (i.e. where the radio emission appears to turn off at the frequencies and sensitivity of LOFAR) then, provided the conditions for generating jets are satisfied, the likelihood of a source being a GSJ is independent of both black hole and stellar mass. This is to be expected since all larger sources must go through the GSJ stage at some point in their evolution. 

Our spectral index analysis indicates that GSJ have a range of ages. While care is needed in interpreting integrated spectral indices, the fact that steep spectrum GSJ tend to have large sizes is consistent with those sources being dominated by older populations of electrons \citep[e.g.][]{Heesen2014ThePopulation}. However, large sources are also present in the sample with flatter spectral indices, which may suggest their average expansion speed is higher than the steeper spectrum sources of similar size. It is possible that some of the steeper spectrum, plausibly older, sources will never grow beyond the GSJ stage. However, all larger radio-galaxies must have been GSJ at some point in their history, and so it is likely that a significant proportion of GSJ do evolve to larger sizes. Detailed population modelling, accounting carefully for selection effects, will be needed to draw stronger conclusions about the relative proportion that will not grow beyond the GSJ phase. 

\subsection{Energetic impact of GSJ}

Though we currently have no direct evidence of the location at which our GSJ are transferring their energy to the external medium, our estimates of the energy supplied show that many are capable of significantly heating the surrounding ISM. This supply of energy to the surrounding environment is therefore capable of restricting the gas cooling rate and reducing the flow of material accreted by the galaxy. This in turn suggests that GSJ are capable of affecting the SFR of their own host galaxy. \citet{Mulchaey2010HotGalaxies} found that low-mass early type galaxies in sparse/isolated environments similar to those of our GSJ have less hot gas than comparable sources in clusters. Our conclusion that GSJ are capable of significantly heating their environments suggests that GSJ may be responsible for moving the gas out to larger radii, causing a decrease in density or even driving a fraction of the ISM from the host galaxy entirely in line with Mulchaey and Jeltema's suggestion that AGN feedback may be responsible for removing this hot gas.

The situation for at least some of our GSJ may be similar to NGC 3801 which, whilst more luminous than our sources, has radio emission about $10\text{ kpc}$ in size and can be considered a GSJ \citep[][]{Heesen2014ThePopulation}. Using X-ray data \citet{Croston2007Shock3801} found that the jets of NGC 3801 were driving shocks into the host ISM. A similar result was found for Centaurus A where the southern radio lobe is driving a shock front into the host galaxy \citep[][]{Croston2009High-energyA}. Similarly Markarian 6 is a Seyfert galaxy with $10\text{ kpc}$ lobes, a luminosity slightly greater than that of our GSJ and radio bubbles of the order $10^{49}\text{ J}$ placing it towards the lower end of the range of energies associated with our GSJ \citep[][]{Mingo2011MarkarianSeyfert, Kharb2006ACycles}. The lobes of Markarian 6 are expanding into the host galaxy creating strong shocks capable of affecting star formation. Finally, Circinus is another Seyfert galaxy where the radio lobes are driving shocks into the host with an energy of about $10^{48}\text{ J}$ \citep[][]{Mingo2012ShocksGalaxy}. These sources suggest that our GSJ may also be capable of producing shocks. Future X-ray studies are needed to confirm if this is the case across the GSJ population.

\section{Summary and Conclusions}
\label{sec:Conclusions}

We have presented a method for efficiently identifying GSJ from within the LoTSS DR1 catalogue. Our main conclusions are:
\begin{itemize}
    \item We have found 195 GSJ with total radio emission no larger than 80 kpc in size; this is the largest sample of intermediate-sized radio galaxies constructed to date.
    \item 9 per cent of the GSJ population are hosted by spiral galaxies, of which two are highly unusual sources generating FRII-like jets.
    \item Our GSJ have luminosities between $3\times10^{22}$ and $1.5\times10^{25}\text{ W Hz}^{-1}$ at 150 MHz.
    \item The host properties of our GSJ show that they are ordinary radio galaxies observed at a stage in their life shortly after the radio emission has expanded beyond the central regions of the host. 
    \item Based on our estimates, about half of our GSJ have internal radio lobe energy within an order of magnitude of the ISM energy. Even ignoring any possible shocks, GSJ are energetically capable of affecting the evolution of their host.
    \item GSJ can occur across a wide range of source ages with many expected to grow into larger sources, making GSJ a key stage in the life cycle of radio galaxies.
\end{itemize}

The LoTSS DR1 covers about 2 per cent of the final survey area. We therefore expect that future releases will uncover an ever-increasing population of GSJ. In the future this will include spectroscopic data for all LOFAR sources greater than $10$ mJy via the WEAVE-LOFAR survey \citep[][]{Smith2016TheSurvey} which will allow for the confirmation of more GSJ sources. Furthermore, we also expect that future LoTSS sub-arcsecond surveys will allow us to detect and study this population at smaller physical sizes where the jets are affecting the inner parts of the host galaxy. This should also allow unambiguous identification of small jets within strongly star-forming galaxies. This increased population and resolution will allow for even more robust studies of this important stage in the life cycle of radio-galaxies. 

Future studies will examine the X-ray properties of a selection of these objects to try and identify the impact GSJ are having upon their environment and to identify the strengths of any shocks produced. This research will allow us to better understand the ability of GSJ to affect their host galaxies and will also allow us to better understand the conditions that determine under what conditions radio-galaxies can grow and develop.

\section*{Acknowledgements}

BW acknowledges a studentship from the UK Science and Technology Facilities Council (STFC). JC and BM acknowledge support from the UK Science and Technology Facilities Council (STFC) under grants ST/R00109X/1 and ST/R000794/1. GG acknowledges the CSIRO research fellowship. MJH acknowledges support from STFC [ST/R000905/1]. HR acknowledges support from the ERC Advanced Investigator programme NewClusters 321271. JS is grateful for support from the UK STFC via grant ST/R000972/1. GJW gratefully acknowledges support of an Emeritus Fellowship from The Leverhulme Trust.

LOFAR, the Low Frequency Array designed and constructed by ASTRON, has facilities in several countries, which are owned by various parties (each with their own funding sources), and that are collectively operated by the International LOFAR Telescope (ILT) foundation under a joint scientific policy. The ILT resources have benefitted from the following recent major funding sources: CNRS-INSU Observatoire de Paris and Universit{\'e} d'Orl{\'e}ans, France; BMBF, MIWF-NRW, MPG, Germany; Science Foundation Ireland (SFI), Department of Business, Enterprise and Innovation (DBEI), Ireland; NWO, The Netherlands; the Science and Technology Facilities Council, UK; Ministry of Science and Higher Education, Poland. Part of this work was carried out on the Dutch national e-infrastructure with the support of the SURF Cooperative through grant e-infra 160022 \& 160152. The LOFAR software and dedicated data reduction packages on \url{https://github.com/apmechev/GRID_LRT} were deployed on the e-infrastructure by the LOFAR e-infragroup, consisting of J.B.R. Oonk (ASTRON \& Leiden Observatory), A.P. Mechev (Leiden Observatory) and T. Shimwell (ASTRON) with support from N. Danezi (SURFsara) and C. Schrijvers (SURFsara). This research has made use of the University of Hertfordshire high-performance computing facility (\url{http://uhhpc.herts.ac.uk}) and the LOFAR-UK computing facility loated at the University of Hertfordshire and supported by STFC [ST/P000096/1].

This publication makes use of data products from the Wide-field Infrared Survey Explorer, which is a joint project of the University of California, Los Angeles, and the Jet Propulsion Laboratory/California Institute of Technology, funded by the National Aeronautics and Space Administration.

Funding for the Sloan Digital Sky Survey IV has been provided by the Alfred P. Sloan Foundation, the U.S. Department of Energy Office of Science, and the Participating Institutions. SDSS acknowledges support and resources from the Center for High-Performance Computing at the University of Utah. The SDSS web site is www.sdss.org.

SDSS is managed by the Astrophysical Research Consortium for the Participating Institutions of the SDSS Collaboration including the Brazilian Participation Group, the Carnegie Institution for Science, Carnegie Mellon University, the Chilean Participation Group, the French Participation Group, Harvard-Smithsonian Center for Astrophysics, Instituto de Astrof\'isica de Canarias, The Johns Hopkins University, Kavli Institute for the Physics and Mathematics of the Universe (IPMU) / University of Tokyo, the Korean Participation Group, Lawrence Berkeley National Laboratory, Leibniz Institut f\"ur Astrophysik Potsdam (AIP), Max-Planck-Institut f\"ur Astronomie (MPIA Heidelberg), Max-Planck-Institut f\"ur Astrophysik (MPA Garching), Max-Planck-Institut f\"ur Extraterrestrische Physik (MPE), National Astronomical Observatories of China, New Mexico State University, New York University, University of Notre Dame, Observat\'orio Nacional / MCTI, The Ohio State University, Pennsylvania State University, Shanghai Astronomical Observatory, United Kingdom Participation Group, Universidad Nacional Aut\'onoma de M\'exico, University of Arizona, University of Colorado Boulder, University of Oxford, University of Portsmouth, University of Utah, University of Virginia, University of Washington, University of Wisconsin, Vanderbilt University, and Yale University.

The Pan-STARRS1 Surveys (PS1) have been made possible through contributions of the Institute for Astronomy, the University of Hawaii, the Pan-STARRS Project Office, the Max-Planck Society and its participating institutes, the Max Planck Institute for Astronomy, Heidelberg and the Max Planck Institute for Extraterrestrial Physics, Garching, The Johns Hopkins University, Durham University, the University of Edinburgh, Queen's University Belfast, the Harvard-Smithsonian Center for Astrophysics, the Las Cumbres Observatory Global Telescope Network Incorporated, the National Central University of Taiwan, the Space Telescope Science Institute, the National Aeronautics and Space Administration under Grant No. NNX08AR22G issued through the Planetary Science Division of the NASA Science Mission Directorate, the National Science Foundation under Grant No. AST-1238877, the University of Maryland, and Eotvos Lorand University (ELTE).

This research made use of the ``K-corrections calculator'' service available at \url{http://kcor.sai.msu.ru/}; Astropy, a community-developed core Python package for Astronomy \citep[][]{Robitaille2013Astropy:Astronomy, Price-Whelan2018ThePackage} available at \url{http://www.astropy.org} and of TOPCAT \citep[][]{Taylor2005TOPCATSoftware}.

This publication uses data generated via the Zooniverse.org platform, development of is which funded by generous support, including a Global Impact Award from Google, and by a grant from the Alfred P. Sloan Foundation.

We also thank the anonymous referee for their useful comments.

\section*{Data Availability}

The data underlying this article are available on the LOFAR Surveys website at \url{https://lofar-surveys.org}, data release 14 of the Sloan Digital Sky Survey at \url{https://skyserver.sdss.org/dr14/en/home.aspx}, the Pan-STARRS1 survey available at \url{https://panstarrs.stsci.edu}, the VLA FIRST Survey at \url{http://sundog.stsci.edu}, the NRAO VLA Sky Survey at \url{https://www.cv.nrao.edu/nvss/} and the Wide-field Infrared Survey Explorer at \url{https://wise2.ipac.caltech.edu/docs/release/allsky/}.




\bibliographystyle{mnras}
\bibliography{Mendeley}


\online{
\newpage
\appendix

\section{Sources in the VS but not the AS}
\label{sec:AppendixPhotoSpecAnalysis}
\label{sec:AppendixAdditionalVSSSources}

In Section~\ref{sec:TheData} of the main text we described the process by which we selected our sources from the LoTSS DR1 catalogue. This led to three sources being included in the VS that were not found in the AS. All three sources are considered individually.

\subsection{ILTJ112543.06+553112.4}
\label{sec:ILTJ112543}

The image of ILTJ112543.06+553112.4 shown in Figure~\ref{fig:ILTJ112543} leaves no doubt that the observed emission is AGN related. The image shows two FRII-like lobes of roughly equivalent flux that are visible in both FIRST and LOFAR. In addition, the sensitivity and resolution of LOFAR shows the jetted radio emission joining these two structures. Whilst the identified host clearly lies closer to the north west lobe such an asymmetry is not unusual in radio galaxies. Additionally, the centre of the identified host lies directly along the line joining the two emission peaks implying that this is indeed the host. However, there does appear to be a small, roughly circular, area of emission bordering the galaxy to the south east. It is unclear what this emission is as it is only slightly more prominent than the remainder of the jetted emission, it could be due to a knot or some other feature within the jet itself. However, it is also possible that this host has been misidentified and that the emission emanates from the core of an unseen host.

To see if the host has been misidentified we re-examined the SDSS survey \citep[][]{Abolfathi2018TheExperiment}; however, no other potential hosts could be found. In particular, there are no sources located at the position of this potential core. The SDSS survey is 95 per cent complete at a limiting magnitude of 21.6, 22.2, 22.2, 21.3 and 20.7 in the u,g,r,i and z-bands respectively so that if this were the core of an unseen host it would have to be fainter than these magnitudes. Similarly the Pan-STARRS survey \citep[][]{Chambers2016TheSurveys} also has no detections within the area of the potential core. The Pan-STARRS survey goes even deeper than SDSS reaching a limiting magnitude of $\sim24$. This effectively rules out the possibility of the radio emission being due to a galaxy within the local universe, although it may still be due to a quasar or galaxy that is located beyond the sensitivity of our instruments.

We must therefore consider the possibility of an unseen, high-redshift host. High redshift radio galaxies do have very high luminosities typically ranging from $L_{150}=10^{26}$ to $10^{30}\text{ W Hz}^{-1}$ \citep[][]{DeBreuck2010THESURVEY,Saxena2019TheRedshifts}. 

The SDSS Quasar survey spectroscopically analyses all quasars with an i-band absolute magnitude less than -20.5 \citep[][]{Paris2018TheRelease}. Ignoring the effects of dust extinction and assuming a flat spectrum a quasar at this magnitude would be visible within SDSS if it had a redshift of $\sim0.4$ or less and visible in Pan-STARRS it it were at a redshift of $\sim1.1$ or less. It is possible that there is an unseen host beyond these redshifts.

Were the observed radio emission located at the SDSS limiting redshift it would have a total linear extent of approximately $450\text{ kpc}$ and a luminosity at $150\text{ MHz}$ of about $10^{26} \text{ W Hz}^{-1}$. At the Pan-STARRS limiting redshift it would have an absolute size of over $650\text{ kpc}$ and a $150\text{ MHz}$ luminosity of about $10^{27} \text{ W Hz}^{-1}$.

Out of the 15,472 LoTSS DR1 sources with sizes measured by LoMorph (see Section~\ref{sec:size-selec} of the main text), 74 ($\sim 0.005$ per cent) are both larger than 650 kpc and have 150 MHz luminosities greater than $10^{27} \text{ W Hz}^{-1}$. Whilst the number of sources with measured redshifts (and hence measured luminosities) within the LoTSS DR1 survey starts decreasing at about $z\sim0.8$, Mingo et al (in prep) has used data from the forthcoming LOFAR Deep Field Survey to show that the distribution of radio source sizes is broadly similar at redshifts up to about $1.5$.

Therefore, an unseen optical host located at the edge of our detection range, generating radio jets of the required size and luminosity would, whilst not unheard of, be rare. At higher redshifts the size and luminosity of the radio emission would have to be even larger making such a source increasingly unlikely. Whilst further observations are required to definitively identify the host galaxy, we believe that the location of the host on the line joining the two emission peaks plus the fact that the potential `core' could easily be due to some anomaly within the jet mean the host has been correctly identified and we are happy to leave this source in the VS.

\begin{figure}
\centering
\includegraphics[width=0.9\columnwidth, trim={0 0 0 0},clip]{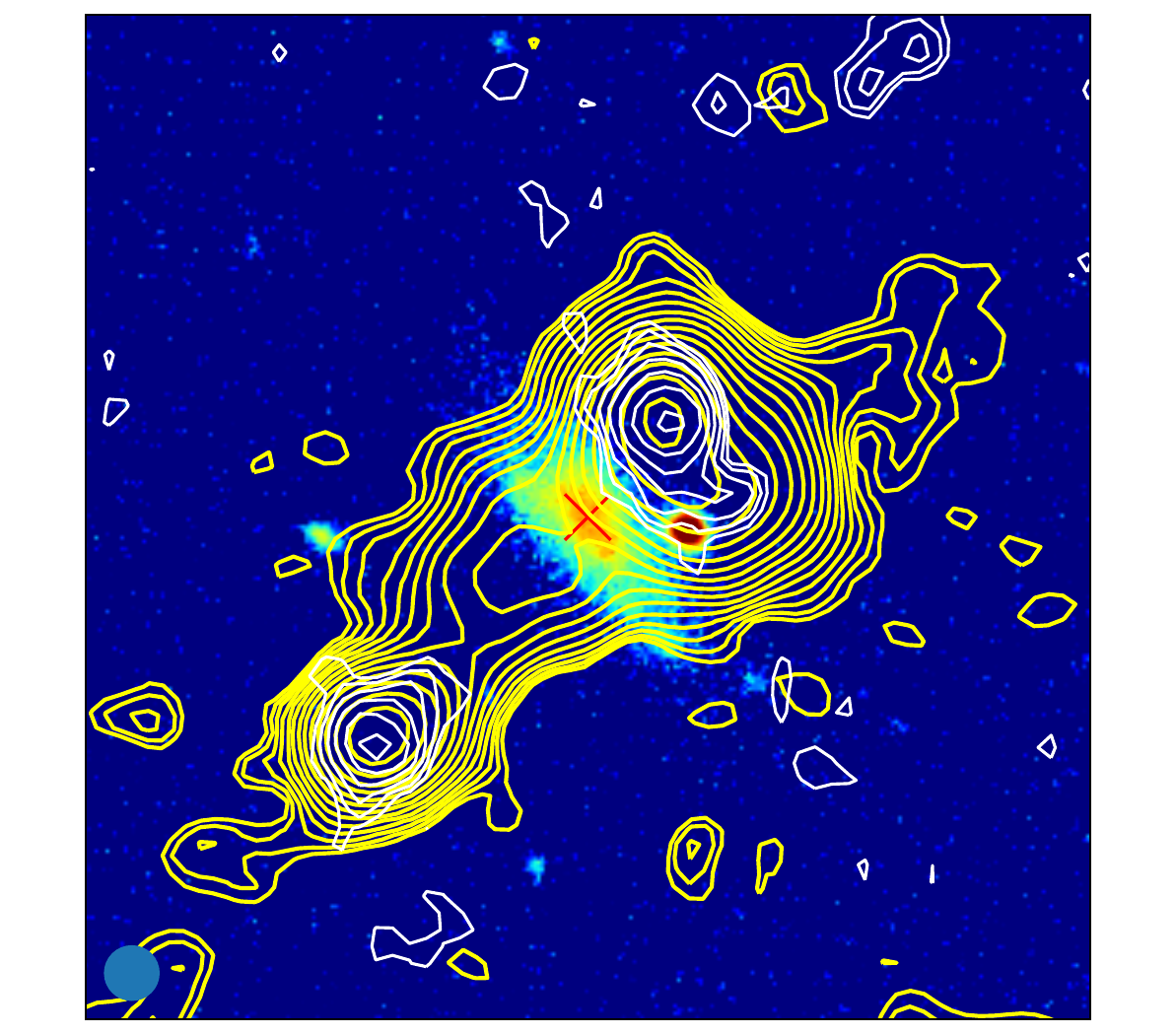}
\caption{Image of ILTJ112543 showing the LOFAR emission in yellow and the FIRST emission in white overlaid on the SDSS r-band optical image. The red cross shows the position of the optical ID. The LOFAR beam is shown in the bottom left.}
\label{fig:ILTJ112543}
\end{figure}

\subsection{ILTJ121847.41+520128.4}
\label{sec:ILTJ121847}

One of the most striking features of ILTJ121847.41+520128.4 is the extreme angle of the lobes which are inclined at an angle of approximately $20^{\circ}$ compared to the major axis of the galaxy (Figure~\ref{fig:DiffPositionAngles}). Whilst jets can exist at any angle with respect to the host, the majority tend to have differences between $40^{\circ}$ and $60^{\circ}$ \citep[][]{Gallimore2006ANuclei}. Even if it does not preclude the jet-related nature of the emission, the angle of $20^{\circ}$ certainly makes this an unusual object.

\begin{figure}
\centering
\includegraphics[width=0.9\columnwidth, trim={0 0 0 0},clip]{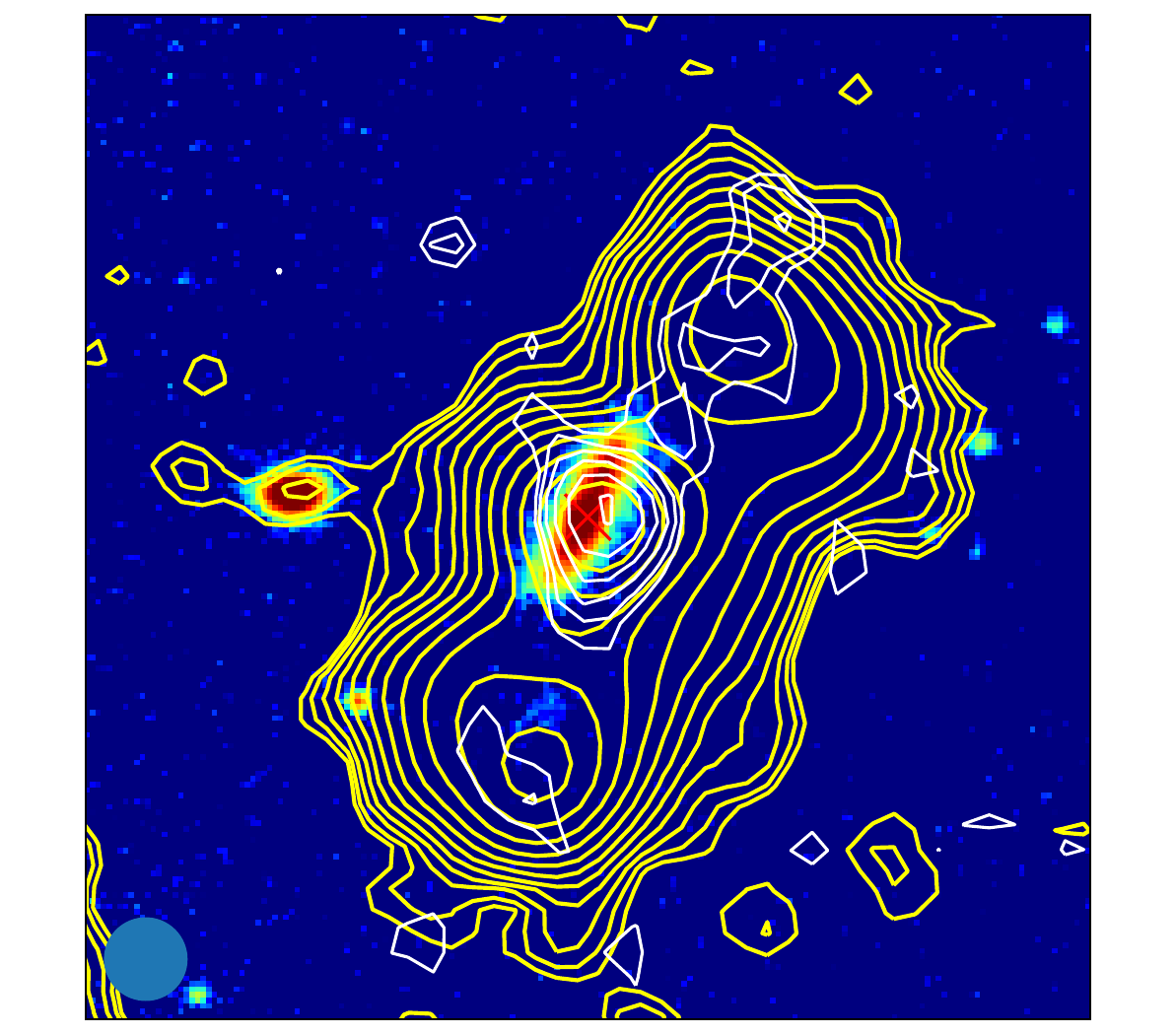}
\caption{Image of ILTJ121847.41+520128.4 showing the LOFAR emission in yellow and the FIRST emission in white overlaid on the SDSS r-band optical image. The red cross shows the position of the optical ID. The LOFAR beam is shown in the bottom left.}
\label{fig:DiffPositionAngles}
\end{figure}

The WISE colour-colour plot for all the selected sources can be seen in Figure~\ref{fig:WISEcc_ILTJ121847}, with ILTJ121847.41+520128.4 highlighted with a black circle. The lines shown on the plot are those used by \citet{Mingo2016TheFIRST/NVSS} to identify AGN ($W1-W2 \geq 0.5$), Ultraluminous Infrared Galaxies (ULIRGs) ($W2-W3 \geq 3.4$) plus the `typical' demarcation line between spiral and eliptical hosts ($W2-W3 = 1.6$). Using this classification, the location of this source in the bottom right of the plot indicates this galaxy may be a ULIRG. Since the dusty nature of ULIRGs means they are sites of intense star formation the possibility of star driven winds must be considered, although the existence of star formation does not preclude there being a jet as well. In particular since the $W1-W2$ colour is $0.332$ (i.e. less than 0.5) this suggests that the emission from the host is star formation dominated although these classifications are less reliable at lower luminosities \citep[][]{Herpich2016TheView,Mateos2012UsingColours}. It has been shown that due to pressure gradients within the ISM stellar winds are typically aligned with the minor axis \citep[][]{Kharb2016A}, which the radio emission in this source is certainly not. This, the FRII-like morphology of the source and the fact that most ULIRGs also display AGN activity \citep[][]{Somerville2015PhysicalFramework}, means the observed emission is almost certainly jet related. 

\begin{figure}
    \centering
    \includegraphics[width = 0.4\textwidth]{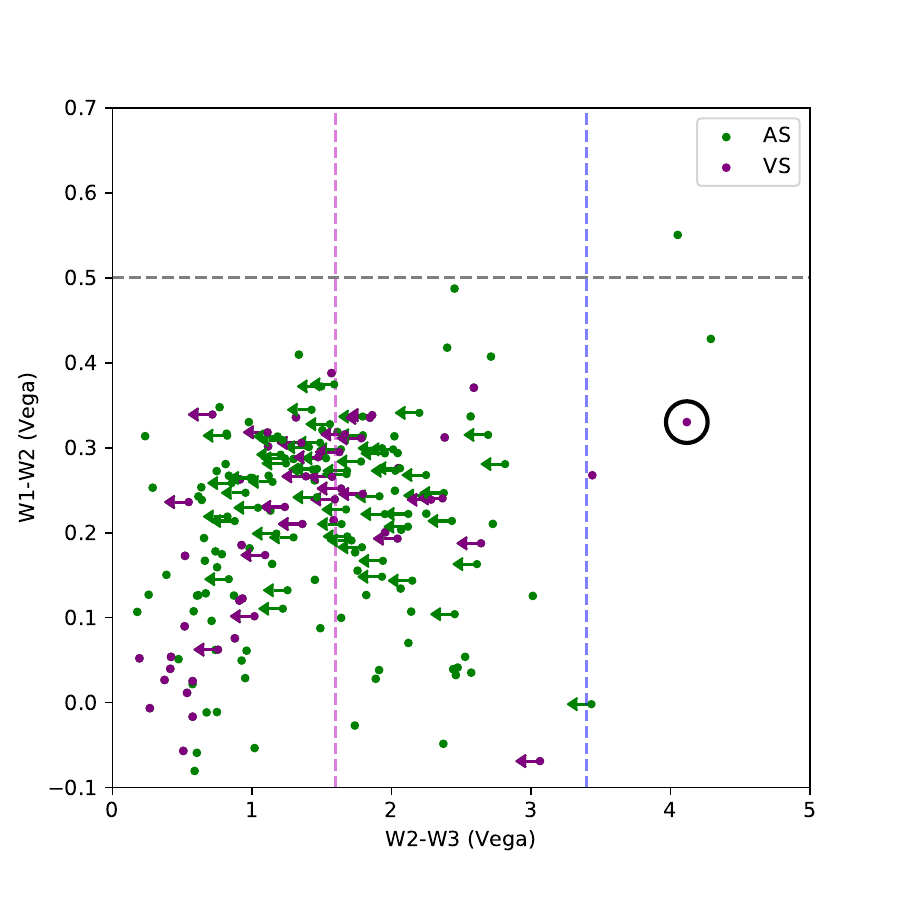}
    \caption{WISE colour colour plot showing the AS (green) and VS (purple) objects with all WISE colours given in the Vega system. The position of ILTJ121847 is circled in black.}
    \label{fig:WISEcc_ILTJ121847}
\end{figure}

The dusty nature of ULIRGs is often explained as being the result of mergers and is expected to provide sufficient fuel to start AGN activity \citep[][]{Wright2010ThePerformance}. Therefore the possibility of merger-induced AGN activity should also be considered. In particular since it is commonly assumed that the orientation of the jet is determined by the black hole's spin \citep[e.g.][]{Tchekhovskoy2011EfficientHole, Gardner2018WhatQuasars}, a binary merger where the spin axis of two supermassive black holes were misaligned could explain extreme jet angles. For example in this case one black hole oriented at approximately $20^{\circ}$ to the major axis of the other galaxy could explain the observed jet angle. This possibility is not precluded by the fact that the host does not show any obvious morphological signs of having undergone a recent merger since AGN activity can be triggered at any time during the merging process. In particular it is anticipated that many low-luminosity AGN will be triggered only after the merger process is complete \citep[][]{Tadhunter2016RadioEvolution}. Whilst ULIRGs are generally associated with major mergers this need not always be the case \citep[][]{Somerville2015PhysicalFramework} plus the diffuse nature of the tell-tale signs of past mergers can often have very low surface brightnesses \citep[][]{Tadhunter2016RadioEvolution} making them difficult to observe, especially in the case of a minor merger where significant disturbance would not be expected \citep[][]{Lotz2010TheMergers}.

It is also possible that the jets are in fact produced by a second galaxy that is obscured from view. Whilst both the FIRST and LOFAR emission do appear very slightly offset from the galaxy center they are still located within the region where an AGN would be located. Although unused as part of their criteria, \citet{Hardcastle2019Radio-loudSources} did identify a luminosity above which hosts can be considered to have a radio-excess resulting from the presence of a radio-loud AGN. This is the black line shown in Figure~\ref{fig:RadioLoudILTJ121847}. Whilst Hardcastle et al. state that this line is unreliable for quiescent galaxies where the star formation rates may be underestimated, the host of ILTJ121847.41+520128.4 is certainly not quiescent and its locus on this plot (shown by the black circle) indicates the presence of an AGN within the observed host, although the fact that the host is a spiral may mean that tests such as this, which have been calibrated using predominantly eliptical hosted galaxies, may be unreliable.

The core has a spectral index of -0.6, which is fairly typical of an AGN \citep[][]{Heesen2014ThePopulation} and although not atypical of star formation \citep[][]{Condon2016EssentialAstronomy}, when combined with the fact that this galaxy also has a radio excess (Figure \ref{fig:RadioLoudILTJ121847}) strongly suggests we are observing the host. The location of the core and lack of any obvious signs of a major merger mean this would have to be occurring directly along our line of sight making this arrangement highly unlikely.

However, despite having properties consistent with being a high excitation source (see Section~\ref{sec:Discussion} of the main text), plotting this source on a BPT diagram (see Figure~\ref{fig:BPT_ILTJ121847}) and using the \citet{Kauffmann2003TheAGN} dividing line suggests that this host does not contain an AGN. The criteria of Kauffmann et al. can misclassify sources with low emission line strengths \citep[for example, see][]{Stasinska2006Semi-empiricalHosts} plus, as above, the spiral nature of the host may mean that this method is unreliable. Therefore, whilst the radio properties suggest the host has been correctly identified, the optical properties do not. It remains possible that the host has been misidentified and that the true host is obscured from view.

\begin{figure}
    \centering
    \includegraphics[width = 0.4\textwidth]{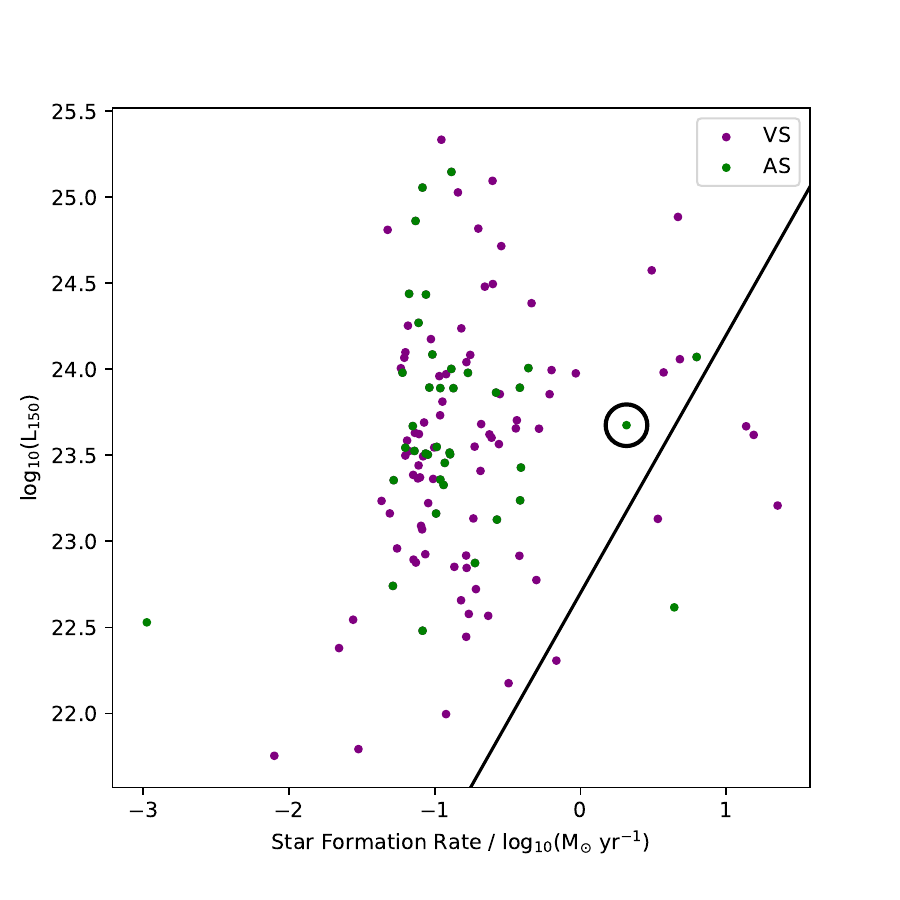}
    \caption{Plot of $L_{150}$ against the Median star formation rate taken from the MPAJHU database. The black line shows the division proposed by \citet{Hardcastle2019Radio-loudSources} for radio-loud / star-forming galaxies. ILTJ121847 is circled in black}
    \label{fig:RadioLoudILTJ121847}
\end{figure}

\begin{figure}
    \centering
    \includegraphics[width=0.4\textwidth]{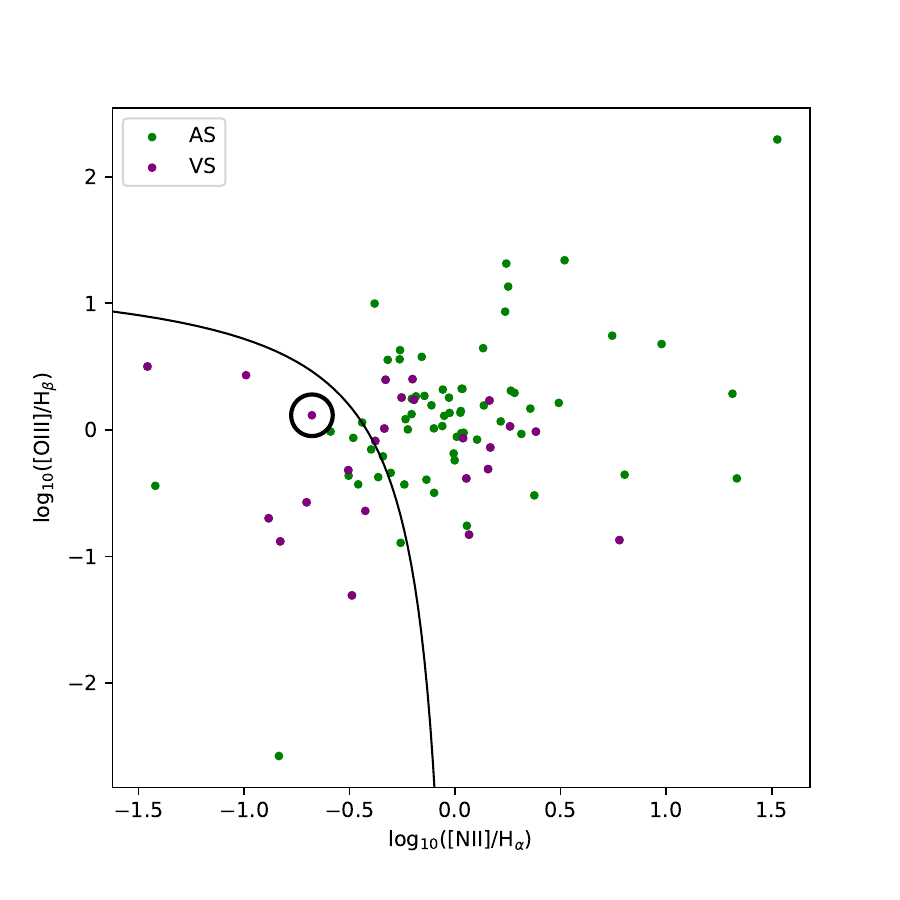}
    \caption{BPT plot for the AS and VS. ILTJ121847 is shown circled in black. The dividing line of \citet{Kauffmann2003TheAGN} separating objects classed as star-forming and AGN is shown in black.}
    \label{fig:BPT_ILTJ121847}
\end{figure}

In summary we conclude that this system is a ULIRG galaxy that is also producing genuine galaxy-scale jets. The fact that this galaxy is a ULIRG does not affect either the optical or X-ray measurements obtained for the galaxy. The angle of the jets also means that our measurements of the jet/lobe radio flux is largely free from any contamination caused by the star-forming regions. We are therefore satisfied that this object should remain in the VS.

\subsection{ILTJ123158.50+462509.9}
\label{sec:ILTJ123158}

ILTJ123158.50+462509.9 is a spectroscopic source with a redshift of $z=0.11$. Morphologically it strongly resembles a jetted source with extended radio emission to both the east and west of the source (see Figure~\ref{fig:ILTJ123158}). Both structures have similar lengths and luminosities suggesting we are looking at the source close to face on. Whilst there is a faint background galaxy that could explain some of the western emission, this is unlikely to be responsible for the rest of the observed emission and no other background sources are detected in either SDSS or PanSTARRS. 

The optically identified host lies in the middle of the radio emission and exhibits a small amount of additional radio flux that could be due to either star formation or a radio core, though we do note that the central emission is slightly off-centre making the star formation explanation slightly more likely though neither possibility can be ruled out. Overall, we are confident the host has been correctly identified.

\begin{figure}
\centering
\includegraphics[width=0.9\columnwidth, trim={0 0 0 0},clip]{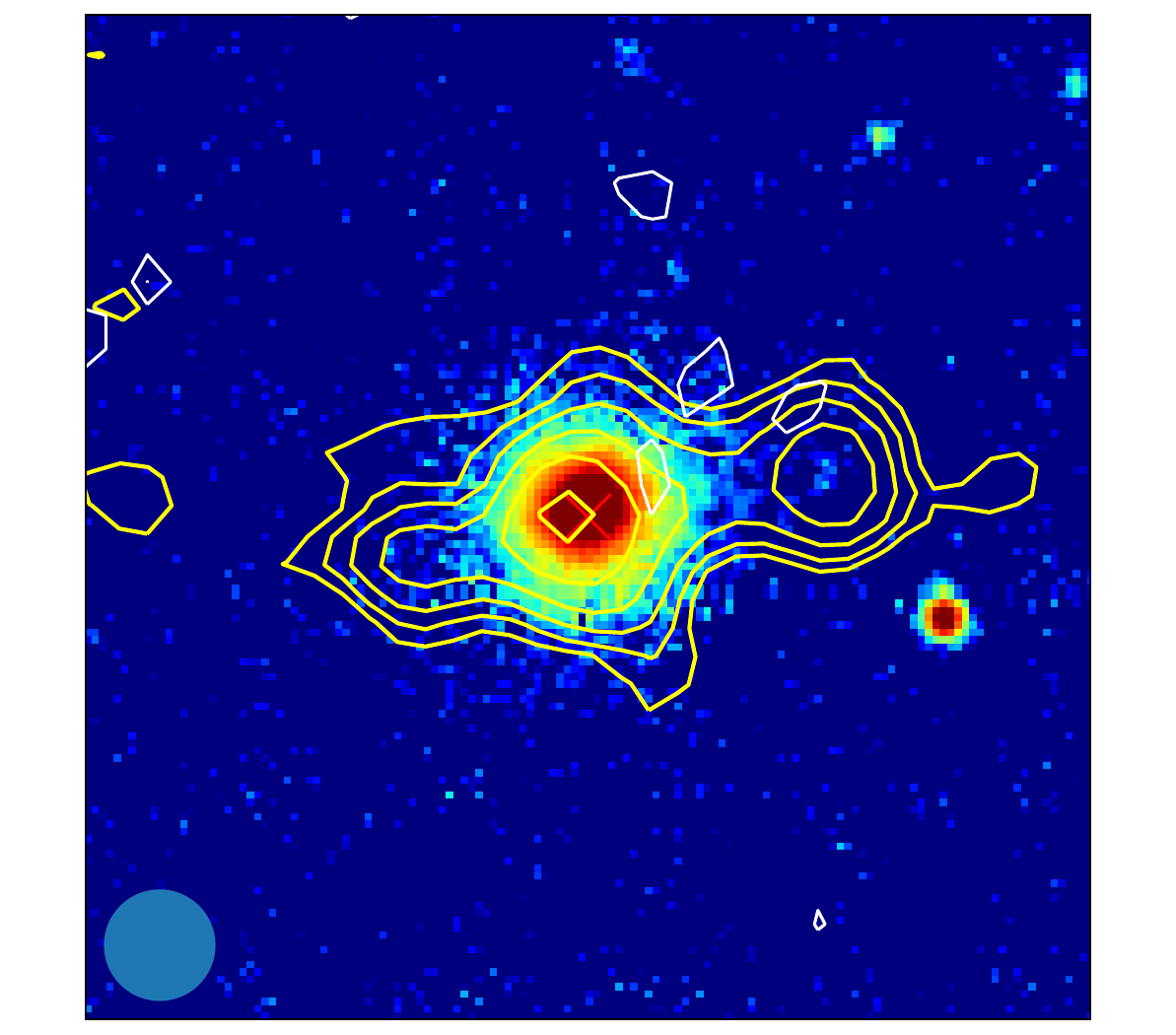}
\caption{Image of ILTJ123158.50+462509.9 showing the LOFAR emission in yellow and the FIRST emission in white overlaid on the SDSS r-band optical image. The red cross shows the position of the optical ID. The LOFAR beam is shown in the bottom left.}
\label{fig:ILTJ123158}
\end{figure}

As a spectroscopic source, ILTJ123158.50+462509.9 was analysed by \citet{Sabater2019TheOn} based on its position in two radio-loudness plots, the BPT diagram and the WISE colour-colour plot. However, all four of their tests suggested this was not an AGN leading Hardcastle et al. to classify this as a star-forming galaxy. This is an anomalous source having one of the lowest radio luminosities in our sample ($L_{150} \sim 4\times10^{22}\text{ W Hz}^{-1}$) but also one of the highest star formation rates ($\sim10^{0.6}\text{ M}_{\odot} yr^{-1}$ according to the MPA-JHU database) and is a borderline ULIRG on the WISE colour-colour plot. Since radio luminosity is expected to increase with higher star formation rates, this may explain why it failed the radio-loudness tests of Hardcastle et al.

The catalogue of Hardcastle et al. aims to be clean rather than complete so that that some genuine AGN are excluded from their sample. As the only elliptically-hosted source in the VS that is not also in the AS this source does therefore suggest that there may be a small sub-sample of GSJ that cannot be detected using traditional methods alone.

\section{Visual Sample}
\label{sec:AppendixVSSSourceImages}

Figure~\ref{fig:VS_Sample} shows a gallery of the sources in the Visual Sample.

\begin{figure*}
\centering
\includegraphics[width=17.7cm]{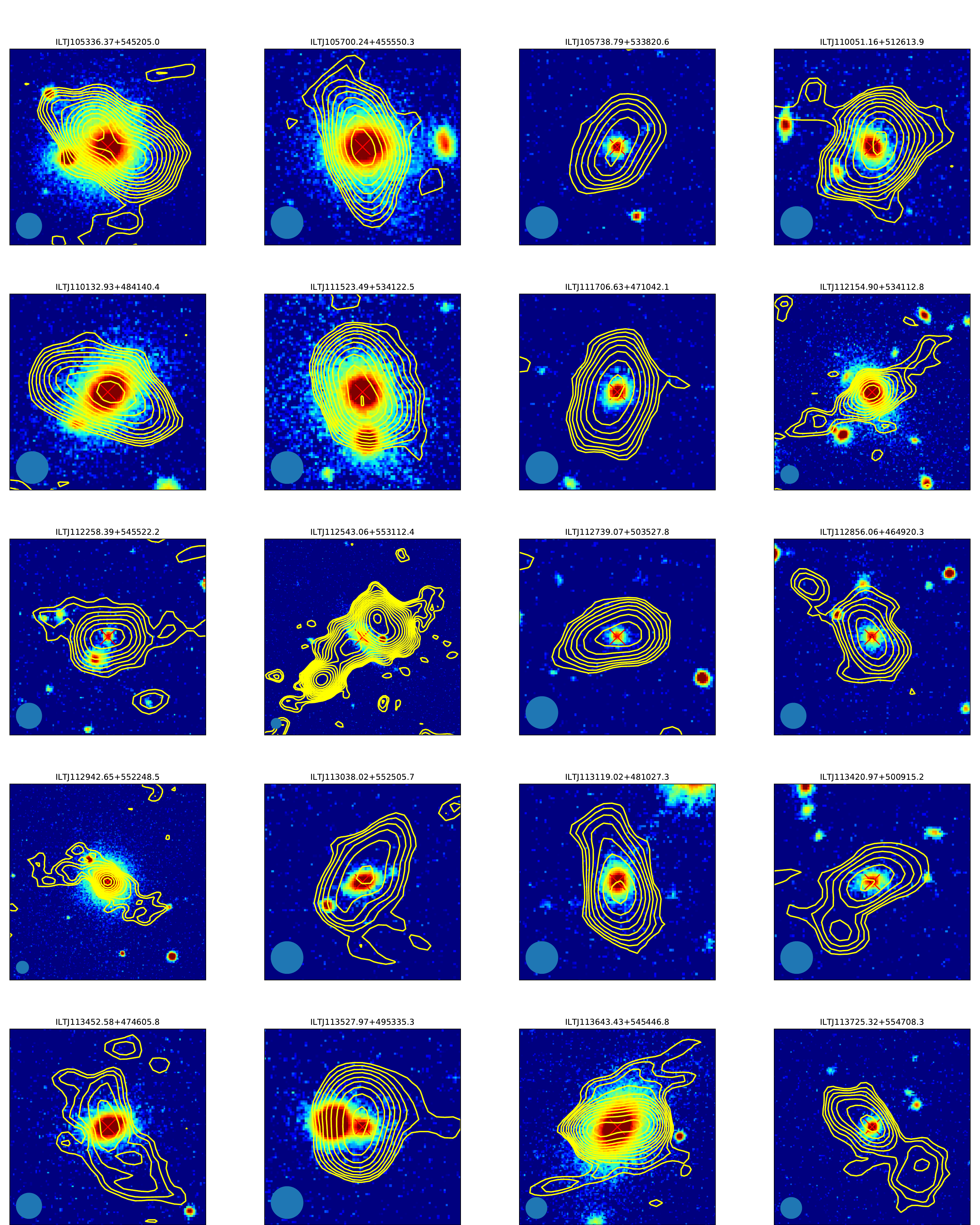}
\caption{Images of all the VS sources. Images show LOFAR contours in yellow overlaid on an r-band SDSS image in colour.}
\label{fig:VS_Sample}
\end{figure*}

\begin{figure*}
\ContinuedFloat
\centering
\includegraphics[width=17.7cm]{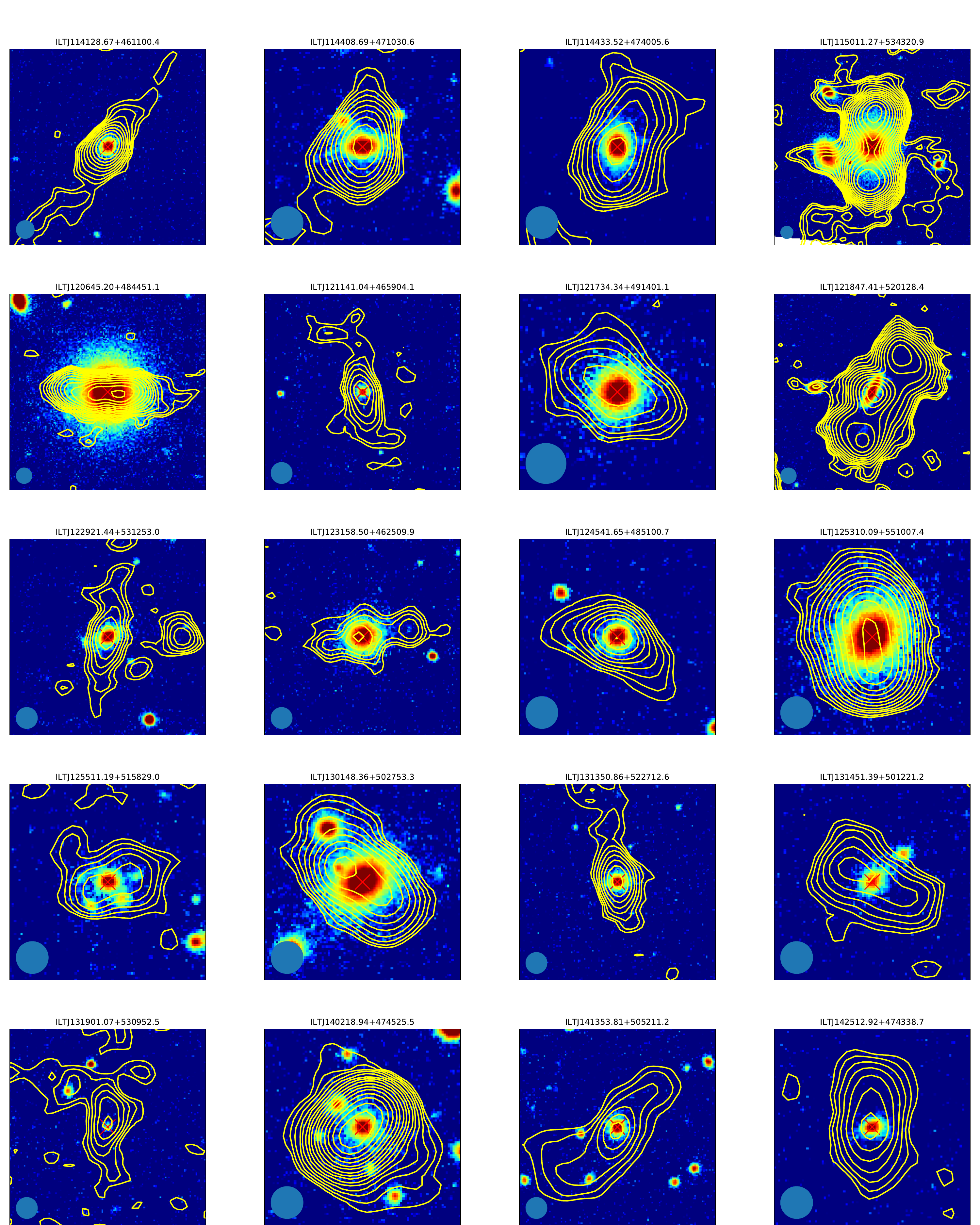}
\caption{Continued}
\end{figure*}

\begin{figure*}
\ContinuedFloat
\centering
\includegraphics[trim=0 720 0 0, clip, width=17.7cm]{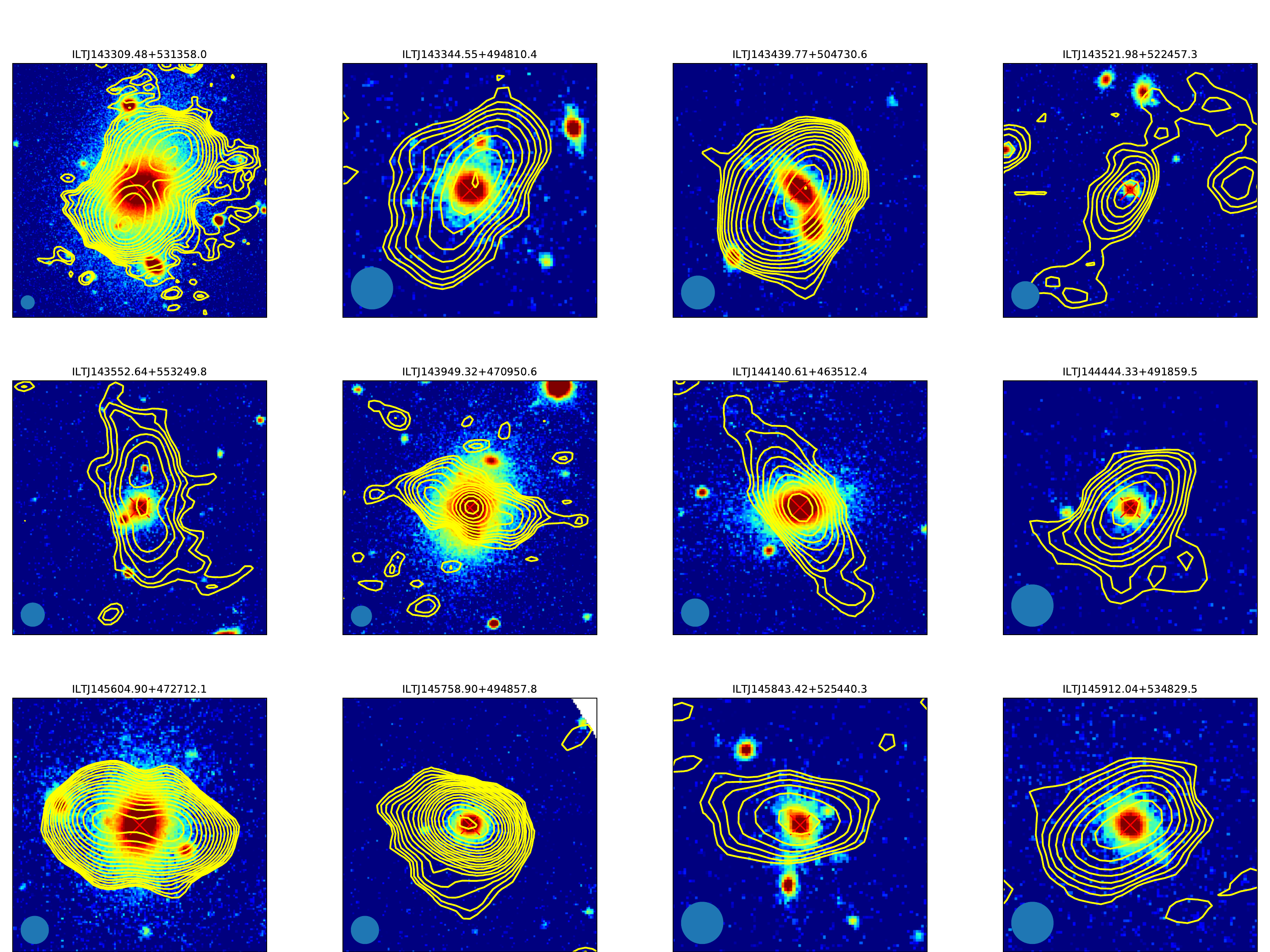}
\caption{Continued}
\end{figure*}
}


\bsp	
\label{lastpage}
\end{document}